\documentstyle[12pt,uartopt]{article}

\newfont{\goth}{cmbxti10 scaled\magstep1}
 \newfont{\gothi}{cmbxti10}

 \newcommand{\smin}{\,\raisebox{0.06em}{${\scriptstyle \in}$}\,}
 \newcommand{\smsubset}{\,\raisebox{0.06em}{${\scriptstyle \subset}$}\,}

 \newcommand{\Real}{\mbox{I \hspace{-0.82em} R}}
 \newcommand{\Complex}{
        \mbox{C \hspace{-1.16em} \raisebox{-0.018em}{\sf l}}\;}
 \newcommand{\gotg}{\mbox{\goth g}}
 \newcommand{\gotig}{\mbox{\gothi g}}
 \newcommand{\oktimes}{\stackrel{\otimes}{,}}
 \newcommand{\smotimes}{\,{\scriptstyle \otimes}\,}

\newcommand{\field}{\mbox{I \hspace{-0.85em} K}}

\newtheorem{theorem}{Theorem}[section]

\newenvironment{proof}{\begin{description}
                       \item[{\small {\bf Proof:}}] \small}{\hfill {\bf
q.e.d..}
                                                            \medskip
                                                            \end{description}}

\begin{document}

\renewcommand{\thefootnote}{\fnsymbol{footnote}}

\title{The Lie-Poisson Structure of Integrable Classical Non-Linear Sigma
       Models}

\author{M. Forger$\,^1$~,~~
        M. Bordemann$\,^1$ \footnotemark[1]~,~~
        J. Laartz$\,^2$ \footnotemark[2]~,~~
        U. Sch\"aper$\,^1$}
 \footnotetext[1]{Supported by the Deutsche
                  Forschungsgemeinschaft, Contract No.~Ro 864/1-1}
 \footnotetext[2]{On leave of absence from Fakult\"at f\"ur Physik
                  der Universit\"at Freiburg, \\
                  \hspace*{6.5mm}Supported by the Studienstiftung des
                  Deutschen Volkes}

 \date{\normalsize
       $^1\,$ Fakult\"at f\"ur Physik der Universit\"at Freiburg, \\
       Hermann-Herder-Str.~3, D-7800 Freiburg / FRG \\[0.2cm]
       $^2\,$ Department of Mathematics, Harvard University, \\
       1 Oxford Street, Cambridge MA 02138 / USA}

\maketitle
\begin{abstract}
\noindent
The canonical structure of classical non-linear sigma models on Riemannian
symmetric spaces, which constitute the most general class of classical
non-linear sigma models known to be integrable, is shown to be governed by
a fundamental Poisson bracket relation that fits into the $r$-$s$-matrix
formalism for non-ultralocal integrable models first discussed by Maillet.
The matrices $r$ and $s$ are computed explicitly and, being field dependent,
satisfy fundamental Poisson bracket relations of their own, which can be
expressed in terms of a new numerical matrix~$c$. It is proposed that
all these Poisson brackets taken together are representation conditions
for a new kind of algebra which, for this class of models, replaces the
classical Yang-Baxter algebra governing the canonical structure of ultralocal
models. The Poisson brackets for the transition matrices are also computed,
and the notorious regularization problem associated with the definition of
the Poisson brackets for the monodromy matrices is discussed.
\end{abstract}

\vfill

 \begin{flushright}
 \parbox{12em}
 { \begin{center}
 University of Freiburg \\
 THEP 91/11 \\
 December 1991\\
 ~ \\
 Harvard University
 \end{center} }
 \end{flushright}

\newpage
\renewcommand{\thefootnote}{\arabic{footnote}}

\setcounter{page}{1}
\section{Introduction}

During the last two decades, there has been great progress in understanding the
structure of two-dimensional integrable field theories. Within the Hamiltonian
approach \cite{FaTa}, there have emerged new algebraic structures, notably the
concept of Yang-Baxter algebras. These algebras appear, e.g., through the
commutation relations of monodromy matrices when solving the models by the
inverse scattering method.

At the classical level one begins by rewriting the equations of motion as a
zero curvature condition, i.e., as the compatibility condition for a linear
system with spectral parameter (Lax pair). In the Hamiltonian context one
then studies the Poisson brackets between $L$-matrices (the spatial part of
the Lax pair). In the most common cases, this leads to a Lie-Poisson algebra
of the form
\begin{equation}
 \{ L(x,\lambda) \oktimes L(y,\mu) \}~~
 =~~\big[ \, r(\lambda,\mu) \; , \;
          L(x,\lambda) \otimes 1 \, + \, 1 \otimes L(y,\mu) \, \big] \;
    \delta(x-y)~~~,                                             \label{eq:FPB1}
\end{equation}
with an antisymmetric $r$-matrix which is numerical, i.e., field independent,
and which obeys the classical Yang-Baxter equation
\begin{equation}
 [ \, r_{12}(\lambda,\mu) , r_{13}(\lambda,\nu) \, ] \, + \,
 [ \, r_{12}(\lambda,\mu) , r_{23}(\mu,\nu) \, ] \, + \,
 [ \, r_{13}(\lambda,\nu) , r_{23}(\mu,\nu) \, ]~~=~~0~~~.      \label{eq:YBE1}
\end{equation}
Theories for which the Poisson brackets between $L$-matrices are of this
form are commonly called ultralocal \cite{FaTa} because the rhs.\ of eqn
(\ref{eq:FPB1}) contains only the delta function $\delta(x-y)$ but not
its derivatives.

An important generalization of the above Lie-Poisson structure to certain
non-ultralocal models, namely those for which the rhs.\ of eqn (\ref{eq:FPB1})
contains, apart from the delta function, its first derivative (i.e., a
classical Schwinger term), has been developed by Maillet \cite{Mai1}.
In his $r$-$s$-matrix approach, eqn (\ref{eq:FPB1}) is replaced
by\footnote{Our $L$ corresponds to $-L$ in ref.~\cite{Mai1}.}
\begin{eqnarray}
\lefteqn{\{ L(x,\lambda) \oktimes L(y,\mu) \}}
                                             \hspace{1.2cm} \nonumber \\[0.2cm]
 &=& {} - \big[ \, r(x,\lambda,\mu) \; , \;
                L(x,\lambda) \otimes 1 \, + \, 1 \otimes L(x,\mu) \, \big] \;
          \delta(x-y)                                       \nonumber \\[0.1cm]
 & & {} + \big[ \, s(x,\lambda,\mu) \; , \;
                L(x,\lambda) \otimes 1 \, - \, 1 \otimes L(x,\mu) \, \big] \;
          \delta(x-y)                                       \nonumber \\[0.1cm]
 & & {} - \big( r(x,\lambda,\mu) + s(x,\lambda,\mu) -
                r(y,\lambda,\mu) + s(y,\lambda,\mu) \big) \;
          \delta^\prime(x-y)~~~,                                \label{eq:FPB2}
\end{eqnarray}
or equivalently,\footnote{The equivalence of eqns (\ref{eq:FPB2}) and
(\ref{eq:FPB3}) follows from the identity (\ref{eq:ID1}) given below.}
\begin{eqnarray}
 \{ L(x,\lambda) \oktimes L(y,\mu) \}
 &=& r^\prime(x,\lambda,\mu) \; \delta(x-y)                 \nonumber \\[0.1cm]
 & & {} - \big[ \, r(x,\lambda,\mu) \; , \;
                L(x,\lambda) \otimes 1 \, + \, 1 \otimes L(x,\mu) \, \big] \;
          \delta(x-y)~~~                                    \nonumber \\[0.1cm]
 & & {} + \big[ \, s(x,\lambda,\mu) \; , \;
                L(x,\lambda) \otimes 1 \, - \, 1 \otimes L(x,\mu) \, \big] \;
          \delta(x-y)~~~                                    \nonumber \\[0.1cm]
 & & {} - \big( s(x,\lambda,\mu) + s(y,\lambda,\mu) \big) \;
          \delta^\prime(x-y)~~~,                                \label{eq:FPB3}
\vspace{0.1cm}
\end{eqnarray}
with an antisymmetric $r$-matrix and a symmetric $s$-matrix which, in general,
depend explicitly on the fields of the theory. Some of the most important
non-ultralocal models, such as the $O(N)$-sigma model \cite{DEM,Mai2},
the principal chiral model \cite{DEM,Mai3} and the complex sine-Gordon model
\cite{Mai1}, are known to fit into this scheme.

In the present paper, we shall demonstrate that non-linear sigma models
defined on Riemannian symmetric spaces, which constitute the general class
of non-linear sigma models known to be integrable \cite{EF,For}, provide
an explicit realization of this structure, with simple formulas expressing
$r$ and $s$ as functions of the basic fields. In particular, the results
of \cite{Mai2,Mai3} are special cases of the one obtained here.

At first sight, the fundamental Poisson brackets (\ref{eq:FPB2}) or
(\ref{eq:FPB3}) are not particularly enlightening. They can however
be brought into a more transparent form, which is also closer to the
$r$-matrix approach used in classical mechanics \cite{BaVi}, by passing
from the $L$-matrix $L(x,\lambda)$ to the corresponding Lax operator
$D(x,\lambda)$, defined as
\begin{equation}
D(x,\lambda)~=~{\partial \over \partial x} \, + \, L(x,\lambda)~~~.
                                                                  \label{eq:LO}
\end{equation}
Namely, the Poisson brackets of the $D$'s are, by definition, the same
as the ones for the $L$'s, but $D$ being a differential operator, the
inhomogeneous classical Schwinger terms on the rhs.\ of eqns (\ref{eq:FPB2})
or (\ref{eq:FPB3}) can be absorbed into commutators (the commutator of a $D$
with a delta function produces, among other things, the derivative of a delta
function). Concretely, it follows directly from the identity
\begin{equation}
 (f(x) - f(y)) \, \delta^\prime(x-y)~=~- \, f^\prime(x) \, \delta(x-y)~~
                                                                 \label{eq:ID1}
\end{equation}
that eqn (\ref{eq:FPB2}), eqn (\ref{eq:FPB3}) and the relation
\begin{eqnarray}
\{ D(x,\lambda) \oktimes D(y,\mu) \}
 &=& - \, \big[ \, r(x,\lambda,\mu) \, \delta(x-y) \; , \;
          D(x,\lambda) \otimes 1 \, + \, 1 \otimes D(y,\mu) \, \big]
                                                            \nonumber \\[0.1cm]
 & & + \, \big[ \, s(x,\lambda,\mu) \, \delta(x-y) \; , \;
          D(x,\lambda) \otimes 1 \, - \, 1 \otimes D(y,\mu) \, \big]~~~~
                                                                \label{eq:FPB4}
\end{eqnarray}
are mutually equivalent.

Besides the $L$-matrix $L(x,\lambda)$ and the Lax operator $D(x,\lambda)$,
another object of central importance in the theory of integrable systems is
the transition matrix $\, T(x,y,\lambda)$, which is simply the corresponding
parallel transport operator at fixed time,
\begin{equation}
 T(x,y,\lambda)~=~P \exp \int_y^x dz~L(z,\lambda)~~~.
\end{equation}
As argued, e.g., in ref.~\cite{Mai1}, eqn (\ref{eq:FPB2}) or (\ref{eq:FPB3})
implies the following basic Poisson brackets between the transition matrices:
\goodbreak
\begin{eqnarray}
\lefteqn{\{ T(x,y,\lambda) \oktimes T(u,v,\mu) \}}
                                             \hspace{0.3cm} \nonumber \\[0.2cm]
 &=& +~\epsilon(x-y) \, \chi(z\,;x,y)                              \nonumber \\
 & & \phantom{+~}
     (T(x,z,\lambda) \otimes T(u,z,\mu)) \; (r+s)(z,\lambda,\mu) \;
     (T(z,y,\lambda) \otimes T(z,v,\mu)) \; \Big|_{z=v}^{z=u}
                                                            \nonumber \\[0.1cm]
 & & +~\epsilon(u-v) \, \chi(z\,;u,v)                              \nonumber \\
 & & \phantom{+~}
     (T(x,z,\lambda) \otimes T(u,z,\mu)) \; (r-s)(z,\lambda,\mu) \;
     (T(z,y,\lambda) \otimes T(z,v,\mu)) \; \Big|_{z=y}^{z=x}~.~~~~~~
                                                                \label{eq:FPB5}
\end{eqnarray}
Here $\epsilon$ is the usual sign function,
\begin{equation}
 \epsilon(x-y)~~=~~\left\{ \begin{array}{ccc}
                            + 1 & {\rm for} & x > y \\
                              0 & {\rm for} & x = y \\
                            - 1 & {\rm for} & x < y
                           \end{array} \right.~~~,               \label{eq:EPS}
\end{equation}
and $\, \chi(.\,;x,y) \,$ is the characteristic function of the interval
between $x$ and $y$. These Poisson brackets are well-defined when considered as
distributions in the respective variables and are continuous functions except
at those points where two of their arguments $\, x,y,u,v \,$ coincide: there,
they exhibit finite jumps proportional to $s$. Due to these discontinuities,
the definition of Poisson brackets between transition matrices for coinciding
or adjacent intervals, and in particular between monodromy matrices, is not
completely straightforward (as would be the case for ultralocal models) but
requires some kind of regularization. As a result, there arises the problem
of finding a regularization scheme such that the fundamental algebraic
properties of Poisson brackets, viz.\ the derivation rule and the Jacobi
identity, remain valid for their regularized counterparts. A priori,
it is not even clear whether such a regularization scheme exists at all,
and indeed it has been suggested in the earlier literature that this is
not the case \cite{DEM}. Later, a prescription which amounts to a ``total
symmetrization over all possible boundary values'' was found to meet these
requirements \cite{Mai1}, but it is a {\em multi-step regularization} in
the sense that regularization of multiple Poisson brackets with a given
number of factors cannot be reduced to repeated regularization of multiple
Poisson brackets with a smaller number of factors.

\medskip
We conclude this introduction with a brief description of the contents of
this paper.

In Sect.~2 we briefly review the current algebra of integrable non-linear
sigma models which has been derived in a previous paper \cite{FLS} and further
analysed in \cite{La}. There
it has been shown that the algebra closes provided that, in addition to the
components $j_\mu$ of the Noether current, a new composite field $j$ is
introduced.

In Sect.~3, we use this result to prove that the Poisson brackets between
the $L$-matrices are indeed of the form (\ref{eq:FPB2}), with matrices $r$
and $s$ which we determine explicitly as functions of the new field $j$.
We also show how the Poisson brackets (\ref{eq:FPB5}) for the transition
matrices can be derived from the Poisson brackets (\ref{eq:FPB2}) or
(\ref{eq:FPB3}) for the $L$-matrices, and we comment on the possibility
of defining regularized Poisson brackets between transition matrices for
coinciding or adjacent intervals, and in particular between monodromy
matrices, by interpreting (\ref{eq:FPB5}) as an equation for functions
rather than distributions, which requires assigning specific values to
the characteristic function $\chi$ at its points of discontinuity.

In Sect.~4, we complete our analysis by computing the Poisson brackets of the
$L$-matrices with the matrices $r$ and $s$, which are non-trivial because $r$
and $s$ are field dependent. As it turns out, these Poisson brackets can all
be expressed with the help of a new matrix $c$ which is numerical, i.e.,
field independent. In fact, the Lax operators $D$ and an appropriate linear
combination $d$ of $r$ and $s$ close to a ``quadratic algebra'', i.e., their
Poisson brackets are linear combinations of terms which are at most quadratic
in $D$ and $d$. Similarly, the transition matrices $T$ and the same linear
combination $d$ of $r$ and $s$ close to an algebra -- at least when all
expressions are considered as distributions in the respective variables.
Finally, we return to the regularization procedure already discussed in the
previous section, which is based on interpreting the basic Poisson brackets
between the $T$'s and $d$'s as equations for functions rather than
distributions
and assigning specific values to the characteristic functions $\chi$ appearing
there at their points of discontinuity -- values that are assumed to be field
independent but are allowed to depend in an arbitrary way on the spectral
parameters involved. We find that a {\em one-step regularization}, which
reduces the computation of regularized multiple Poisson brackets to a repeated
application of the formula for the regularized basic Poisson brackets and the
derivation rule, fails to satisfy the Jacobi identity.

In an appendix, we give, following ref.~\cite{AvTa}, an explicit proof of a
theorem on the classification of a class of asymmetric solutions to the
classical Yang-Baxter equation, related to involutive automorphisms of
semisimple Lie algebras.

Finally, we would like to mention a recent mathematical preprint \cite{BFPP}
where ``sufficiently generic'' harmonic maps from the two-torus into Riemannian
symmetric spaces are classified. The $r$-matrix arising there is however field
independent and seems to play a different role than the $r$-matrix here.

\pagebreak

\section{Current Algebra for Integrable Non-Linear Sigma Models}

We begin by briefly reviewing the results on the current algebra structure
of general non-linear sigma models derived in a previous paper \cite{FLS}.
For simplicity, we restrict ourselves to the class of integrable non-linear
sigma models, which are precisely those defined on Riemannian symmetric spaces
$\; M = G/H \,$, and we shall use the gauge dependent formulation developed in
\cite{EF,For} where the basic field $\varphi$ with values in $M$ is (at least
locally) represented in terms of a field $g$ with values in $G$, determined
modulo a field $h$ with values in $H$ by the condition that $~\varphi = gH \,$.
Technically, we require that $M$ is the quotient space of some (connected)
Lie group $G$, with Lie algebra $\mbox{\goth g}\,$, modulo some compact
subgroup $\; H \smsubset G \,$, with Lie algebra $\; \mbox{\goth h} \smsubset
\mbox{\goth g} \,$, and that there exists an $\mbox{Ad}(H)$-invariant subspace
$\mbox{\goth m}\,$ of $\mbox{\goth g}\,$, with commutation relations
\begin{equation}
 [\mbox{\goth h}\, , \mbox{\goth h}\,] \smsubset \mbox{\goth h}~~~,~~~
 [\mbox{\goth h}\, , \mbox{\goth m}\,] \smsubset \mbox{\goth m}~~~,~~~
 [\mbox{\goth m}\, , \mbox{\goth m}\,] \smsubset \mbox{\goth h}~~~,
                                                                 \label{eq:CR1}
\end{equation}
such that $\mbox{\goth g}\,$ is the (vector space) direct sum of
$\mbox{\goth h}\,$ and $\mbox{\goth m}\,$:
\begin{equation}
 \mbox{\goth g}\, = \mbox{\goth h}\, \oplus \mbox{\goth m}~~~.    \label{eq:DD}
\end{equation}
The corresponding projections from $\mbox{\goth g}\,$ onto $\mbox{\goth h}\,$
along $\mbox{\goth m}\,$ and from $\mbox{\goth g}\,$ onto $\mbox{\goth m}\,$
along $\mbox{\goth h}\,$ will be denoted by $\, \pi_{\mbox{\gothi h}} \,$ and
$\, \pi_{\mbox{\gothi m}} \,$, respectively. Moreover, we suppose that the
$\mbox{Ad}(H)$-invariant positive definite inner product $(.\, ,.)$ on
$\mbox{\goth m}\,$, corresponding to the given $G$-invariant Riemannian
metric on $M$, is induced from an $\mbox{Ad}(G)$-invariant non-degenerate
inner product $(.\, ,.)$ on $\mbox{\goth g}\,$, corresponding to a
$G$-biinvariant pseudo-Riemannian metric on $G$, so that the direct
decomposition (\ref{eq:DD}) is orthogonal. (For symmetric spaces of the
compact or non-compact type, this hypothesis involves no loss of generality.)
Then defining the covariant derivative $\, D_\mu g \,$ of $g$ to be
\begin{equation}
 D_\mu g~=~g \, \pi_{\mbox{\gothi m}} (g^{-1} \, \partial_\mu g)~~~,
\end{equation}
the action of the non-linear sigma model on $\; M = G/H \;$ can be written as
\begin{equation}
 S~=~{\textstyle{1\over 2}} \int d^2\! x \;
     (\partial^\mu \varphi , \partial_\mu \varphi)
  ~=~{\textstyle{1\over 2}} \int d^2\! x \; (D^\mu g , D_\mu g)~~~,\label{eq:S}
\end{equation}
The global $G$-invariance of this action leads to a Noether current $j_\mu$
taking values in $\mbox{\goth g}\,$:
\begin{equation}
 j_\mu~=~\mbox{} - D_\mu g \, g^{-1}~~~.                          \label{eq:NC}
\end{equation}
As usual, the equations of motion imply (and in the models considered here are
in fact equivalent to) current conservation
\begin{equation}
 \partial^\mu j_\mu~=~0~~~.                                       \label{eq:CC}
\end{equation}
In addition, the commutation relations (\ref{eq:CR1}) guarantee that $2j_\mu$
also satisfies the zero curvature condition, i.e.,
\begin{equation}
 \partial_\mu j_\nu - \partial_\nu j_\mu + 2 \, [j_\mu,j_\nu]~=~0~~~.
                                                                 \label{eq:ZC1}
\end{equation}
The other composite field which will be of central importance for all
that follows is the scalar field $j$ taking values in the space
$~L(\mbox{\goth g}\,) \cong \mbox{\goth g} \, \otimes \mbox{\goth g}^\ast~$
of linear transformations on $\mbox{\goth g}\,$ defined as
\begin{equation}
 j~=~\mbox{Ad}(g) \, \pi_{\mbox{\gothi m}} \, \mbox{Ad}(g)^{-1}~~~.
                                                                  \label{eq:AF}
\end{equation}
It should be noted that the fields $j_\mu$ and $j$ are not independent.
Thus for example, we have the algebraic identity
\begin{equation}
 \mbox{ad}(j_\mu)~=~[ \, j \, , \, \mbox{ad}(j_\mu) \, ]_+
                 ~=~j \, \mbox{ad}(j_\mu) \, + \, \mbox{ad}(j_\mu) \, j~~~,
                                                                 \label{eq:ID2}
\end{equation}
as well as an identity expressing the derivatives of $j$ in terms of $j$ and
$j_\mu$:
\begin{equation}
 \partial_\mu j~=~[ \, j \, , \, \mbox{ad}(j_\mu) \, ]
               ~=~j \, \mbox{ad}(j_\mu) \, - \, \mbox{ad}(j_\mu) \, j~~~.
                                                                 \label{eq:ID3}
\end{equation}
The main motivation for introducing the field $j$ is that it is precisely the
additional ingredient needed to write down the current algebra
in closed form. To do so, it is convenient to introduce a basis $(T_a)$
of $\mbox{\goth g}\,$, with structure constants $f_{ab}^c$ defined by
\mbox{$[T_a,T_b] = f_{ab}^c T_c~$} and metric coefficents $\eta_{ab}$
defined by \mbox{$\; (T_a,T_b) = \eta_{ab} \,$}, together with the
corresponding dual basis $(T^a)$ of $\gotg^\ast$, and to expand $j_\mu$
and $j$ into components:\footnote{According to the original definition,
valid for arbitrary Riemannian manifolds, $j_\mu$ should be considered
as taking its values in the dual space $\gotig^\ast$ of $\gotig\,$ while $j$
should be considered as taking its values in the symmetric tensor product of
$\gotig^\ast$ with itself. The point of view taken above only emerges after
identifying $\gotig\,$ with $\gotig^\ast$ by means of the invariant scalar
product $(.\, ,.)$ on $\gotig\,$, or in component language, by using the
corresponding metric coefficients to pull up and down Lie algebra indices.}
\begin{equation}
 j_\mu~=~j_{\mu,a} \, T^a~~~,~~~j~=~j_{ab} \; T^a\otimes T^b~~~.
                                                               \label{eq:COMP1}
\end{equation}
With this notation, the current algebra (at fixed time $t$) takes the form
\cite{FLS}
\begin{eqnarray}
 \{ j_{0,a}(x) , j_{0,b}(y) \}
 &=& - \, f_{ab}^c \, j_{0,c}(x) \, \delta(x-y)~~~,      \label{eq:CA1} \\[1mm]
 \{ j_{0,a}(x) , j_{1,b}(y) \}
 &=& - \, f_{ab}^c \, j_{1,c}(x) \, \delta(x-y) \,
     + \, j_{ab}(y) \, \delta^\prime(x-y)~~~,            \label{eq:CA2} \\[1mm]
 \{ j_{1,a}(x) , j_{1,b}(y) \} &=& \, 0~~~,              \label{eq:CA3} \\[2mm]
 \{ j_{0,a}(x) , j_{bc}(y) \}
 &=& - \left( f_{ab}^d \, j_{cd}(x) + f_{ac}^d \, j_{bd}(x) \right)
                                    \delta(x-y)~~~,      \label{eq:CA4} \\[1mm]
 \{ j_{1,a}(x) , j_{bc}(y) \} &=& \, 0~~~,               \label{eq:CA5} \\[2mm]
 \{ j_{ab}(x) , j_{cd}(y) \} &=& \, 0~~~.                \label{eq:CA6}
\end{eqnarray}

\section{Fundamental Poisson Brackets}

Our goal in this section is to compute, for the models under consideration,
Poisson brackets between various quantities which play a central role in the
theory of two-dimensional integrable field theories, such as the $L$-matrix
$L(x,\lambda)$ or the transition matrix $\, T(x,y,\lambda) \,$ and the
monodromy matrix $T(\lambda)$. Our starting point will of course be the
current algebra (\ref{eq:CA1})-(\ref{eq:CA6}), which governs the canonical
structure of these models.

According to the commonly accepted point of view (see, e.g., ref.~\cite{FaTa}),
integrability of a classical two-dimensional field theory is expressed through
the possibility to rewrite its equations of motion as a zero curvature
condition
\begin{equation}
 \partial_\mu L_\nu - \partial_\nu L_\mu + [L_\mu,L_\nu]~=~0~~~,
                                                                 \label{eq:ZC2}
\end{equation}
or equivalently, as the condition of commutativity
\begin{equation}
 [D_\mu,D_\nu]~=~0                                               \label{eq:ZC3}
\end{equation}
for the covariant derivatives
\vspace{0.2cm}
\begin{equation}
 D_\mu~=~\partial_\mu + L_\mu~~~.                                 \label{eq:CD}
\vspace{0.2cm}
\end{equation}
This is the compatibility condition for the linear system (Lax pair)
\begin{equation}
 \partial_\mu U~=~U \, L_\mu~~~.                                 \label{eq:LS1}
\end{equation}
Here $U$ and $L_\mu$ are functions on two-dimensional space-time taking
values in an appropriate Lie group $G$ and in the corresponding Lie algebra
$\mbox{\goth g}\,$, respectively, and depending on an additional spectral
parameter $\lambda$. In the present case, $L_\mu$ is simply an appropriate
$\lambda$-dependent linear combination of the current $j_\mu$ and its (Hodge)
dual $\, \epsilon_{\mu\nu} j^\nu$, namely\footnote{Our convention for the
spectral parameter follows that of \cite[p.~312]{FaTa} and \cite{Mai2}; it is
related to the spectral parameter used in \cite{DEM} and \cite{Mai1} by the
inversion $\, \lambda \rightarrow \lambda^{-1} \,$ and to the spectral
parameter $\gamma$ employed in \cite{EF,For} by the M\"obius transformation
$~\lambda \, \rightarrow \, \gamma = (\lambda\!-\!1)/(\lambda\!+\!1) \,$.
Our convention for the $\epsilon$-tensor is $~\epsilon_{01} = -1 \,$.}
\begin{equation}
 L_\mu~=~{2 \over 1\!-\!\lambda^2} \left( j_\mu + \lambda
                      \epsilon_{\mu\nu} j^\nu \right)~~~.        \label{eq:LM1}
\end{equation}
Indeed, with this choice, the $\lambda$-dependent zero curvature condition
(\ref{eq:ZC2}) is equivalent to both current conservation (\ref{eq:CC}) and
the $\lambda$-independent zero curvature condition (\ref{eq:ZC1}).

Next, recall that the $L$-matrix $L$ is defined to be the spatial component
of the flat vector potential $L_\mu$,
\begin{equation}
 L(x,\lambda)~=~{2 \over 1\!-\!\lambda^2}
                \left( j_1(x) + \lambda j_0(x) \right)~~~,       \label{eq:LM2}
\end{equation}
while the transition matrix $\, T(x,y,\lambda) \,$ is simply the parallel
transport operator from $y$ to $x$ (at fixed time $t$) associated with the
flat vector potential $L_\mu$, so in terms of the solutions $U$ of the linear
system (\ref{eq:LS1}), it is given by
\begin{equation}
 T(x,y,\lambda)~=~U(x,\lambda)^{-1} \, U(y,\lambda)~~~.          \label{eq:TM1}
\end{equation}
Therefore it obeys the differential equations
\begin{equation}
 {\partial \over \partial x} \, T(x,y,\lambda)
 ~=~- \, L(x,\lambda) \, T(x,y,\lambda)                          \label{eq:TM2}
\vspace{-0.2cm}
\end{equation}
and
\begin{equation}
 {\partial \over \partial y} \, T(x,y,\lambda)
 ~=~+ \, T(x,y,\lambda) \, L(y,\lambda)                          \label{eq:TM3}
\vspace{0.2cm}
\end{equation}
with initial condition
\vspace{0.2cm}
\begin{equation}
 T(x,y,\lambda) \, \Big|_{x=y}~=~1~~~,                           \label{eq:TM4}
\end{equation}
as well as the composition rule
\begin{equation}
 T(x,y,\lambda) \, T(y,z,\lambda)~=~T(x,z,\lambda)~~~,           \label{eq:TM5}
\end{equation}
which, together with (\ref{eq:TM4}), leads to the inversion formula
\begin{equation}
 T(x,y,\lambda)^{-1}~=~T(y,x,\lambda)~~~.                        \label{eq:TM6}
\end{equation}
Moreover, under standard boundary conditions on the fields at spatial infinity
($\varphi$ approaches a given point in $M$ and all of its derivatives vanish
sufficiently rapidly), the limit
\begin{equation}
  T(\lambda)~=~\lim_{{x \to + \infty} \atop {y \to - \infty}} T(x,y,\lambda)~~
                                                                 \label{eq:MM1}
\end{equation}
exists: this is the monodromy matrix.

Turning to the actual calculation of Poisson brackets, we first expand $L$ into
components
\begin{equation}
 L~=~L_a \, T^a~~~,                                            \label{eq:COMP2}
\end{equation}
cf.\ (\ref{eq:COMP1}). In these terms, the result follows directly from
combining the current algebra (\ref{eq:CA1})--(\ref{eq:CA6}) with the
definition (\ref{eq:LM2}) of $L$:
\begin{eqnarray}
 \{ L_a(x,\lambda) \, , \, L_b(y,\mu) \}
 &\!=\!& {2 f_{ab}^c \over \lambda\!-\!\mu} \,
         \Bigl( {\mu^2 \over 1\!-\!\mu^2} \, L_c(x,\lambda) \, - \,
                {\lambda^2 \over 1\!-\!\lambda^2} \, L_c(x,\mu) \Bigr) \;
         \delta(x-y)                                        \nonumber \\[0.1cm]
 &\!   & + \; {4 \over (1\!-\!\lambda^2)(1\!-\!\mu^2)} \,
         \Bigl( \mu j_{ab}(x) + \lambda j_{ab}(y) \Bigr) \;
         \delta^\prime(x-y)~,~~~~~~                   \label{eq:PB1a} \\[0.2cm]
 \{ L_a(x,\lambda) \, , \, j_{bc}(y) \}
 &\!=\!& - \, {2\lambda \over 1\!-\!\lambda^2} \,
         \Bigl( f_{ab}^d \, j_{cd}(x) + f_{ac}^d \, j_{bd}(x) \Bigr) \;
         \delta(x-y)~.~~~~~~                          \label{eq:PB2a} \\[0.2cm]
 \{ j_{ab}(x) \, , \, j_{cd}(y) \} &\!=\!& 0~.~~~~~~            \label{eq:PB3a}
\end{eqnarray}

The next step will be to rewrite the first of these equations in terms of the
usual tensor notation \cite{FaTa} and to show that it can in fact be cast into
the form (\ref{eq:FPB2}) or (\ref{eq:FPB3}) given by Maillet \cite{Mai1}.
To do so, we embed $\gotg\,$ into its universal enveloping algebra $U(\gotg\,)$
and consider the fields $L$ and $j$ as taking values in $U(\gotg\,)$ and in
$\, U(\gotg\,) \otimes U(\gotg\,)$, respectively. Similarly, the Casimir
tensor $C$, defined by
\begin{equation}
 C~=~\eta^{ab} \; T_a\otimes T_b~~~,
\end{equation}
is viewed as an element of $\, U(\gotg\,) \otimes U(\gotg\,)$. Then we have,
for example,
\begin{eqnarray}
 [ \, C \, , \, T_c \otimes 1 \, ]
 ~=~+ \, f^{ab}_{\phantom{ab} c} \; T_a \otimes T_b &,&
 [ \, C \, , \, 1 \otimes T_c \, ]
 ~=~- \, f^{ab}_{\phantom{ab} c} \; T_a \otimes T_b~~~,         \label{eq:CR2}
\end{eqnarray}
while
\begin{eqnarray}
 [ \, j \, , \, T_c \otimes 1 \, ]
 ~=~- \, f_{cd}^a \, j^{bd} \; T_a \otimes T_b &,&
 [ \, j \, , \, 1 \otimes T_c \, ]
 ~=~- \, f_{cd}^b \, j^{ad} \; T_a \otimes T_b~~~,              \label{eq:CR3}
\end{eqnarray}
so that the identities (\ref{eq:ID2}) and (\ref{eq:ID3}) become
\begin{equation}
 [ \, C \, , \, j_\mu \otimes 1 \, ]~=~- \,
 [ \, C \, , \, 1 \otimes j_\mu \, ]~~=~~
 [ \, j \; , \; j_\mu \otimes 1 \, - \, 1 \otimes j_\mu \, ]~~~, \label{eq:ID4}
\end{equation}
and
\begin{equation}
 \partial_\mu j~~=~~
 [ \, j \; , \; j_\mu \otimes 1 \, + \, 1 \otimes j_\mu \, ]~~~, \label{eq:ID5}
\end{equation}
respectively; they can be combined into one identity involving $L$ instead of
the $j_\mu$:
\begin{eqnarray}
 {2(\lambda\!-\!\mu) \over (1\!-\!\lambda^2)(1\!-\!\mu^2)} \, j^\prime(x) \!
 &=& \! \left[ \, C - 2j(x) \; , \;
     {\mu \over 1\!-\!\mu^2} \, L(x,\lambda) \otimes 1 \, - \,
     {\lambda \over 1\!-\!\lambda^2} \, 1 \otimes L(x,\mu) \, \right] .~~~~~~~~
                                                                 \label{eq:ID6}
\end{eqnarray}
Therefore, we can rewrite the Poisson bracket relation (\ref{eq:PB1a})
in the form
\begin{eqnarray*}
 \{ L(x,\lambda) \oktimes L(y,\mu) \} \!
 &=& \! \left[ \, {2 \, C \over \lambda\!-\!\mu} \; , \;
         {\mu^2 \over 1\!-\!\mu^2} \, L(x,\lambda) \otimes 1 \, + \,
         {\lambda^2 \over 1\!-\!\lambda^2}  \, 1 \otimes L(x,\mu) \, \right] \,
        \delta(x-y)                                                   \\[0.1cm]
 & & ~~- \; {2(\lambda\!-\!\mu) \over (1\!-\!\lambda^2)(1\!-\!\mu^2)} \,
     \Bigl( j(x) - j(y) \Bigr) \; \delta^\prime(x-y)                  \\[0.1cm]
 & & ~~+ \; {2(\lambda\!+\!\mu) \over (1\!-\!\lambda^2)(1\!-\!\mu^2)} \,
     \Bigl( j(x) + j(y) \Bigr) \; \delta^\prime(x-y)~~~.
\end{eqnarray*}
Using the identities (\ref{eq:ID1}),(\ref{eq:ID6}) and collecting the terms
gives
\begin{eqnarray*}
 \{ L(x,\lambda) \oktimes L(y,\mu) \} \!
 &=& \! \left[ \, {\lambda\!+\!\mu  \over \lambda\!-\!\mu} \, C \; , \;
         {\mu \over 1\!-\!\mu^2} \, L(x,\lambda) \otimes 1 \, + \,
         {\lambda \over 1\!-\!\lambda^2}  \, 1 \otimes L(x,\mu) \, \right] \,
        \delta(x-y)                                                   \\[0.1cm]
 & & - \, \left[ \, 2 \, j(x) \; , \;
          {\mu \over 1\!-\!\mu^2} \, L(x,\lambda) \otimes 1 \, - \,
          {\lambda \over 1\!-\!\lambda^2}  \, 1 \otimes L(x,\mu) \, \right] \,
          \delta(x-y)                                                 \\[0.1cm]
 & & \;\! +~{2(\lambda\!+\!\mu) \over (1\!-\!\lambda^2)(1\!-\!\mu^2)} \,
            \Bigl( j(x) + j(y) \Bigr) \; \delta^\prime(x-y)~~~.
\end{eqnarray*}
Comparing this with the rhs.\ of eqn (\ref{eq:FPB3}) suggests the following
Ansatz for $r$ and $s$:
\[
 r(z,\lambda,\mu)~=~a(\lambda,\mu) \, C + \, b(\lambda,\mu) \, j(z)~~~,
\]
with coefficients $a(\lambda,\mu)$ and $b(\lambda,\mu)$ to be determined, and
\[
 s(z,\lambda,\mu)~=~- \, {2(\lambda\!+\!\mu) \over
                          (1\!-\!\lambda^2)(1\!-\!\mu^2)} \, j(z)~~~.
\]
Then applying the identity (\ref{eq:ID6}) again, eqn (\ref{eq:FPB3}) becomes,
after a short calculation
\begin{eqnarray*}
\lefteqn{\{ L(x,\lambda) \oktimes L(y,\mu) \}}         \hspace{0.3cm} \\[0.2cm]
 &=& - \, \bigg( a(\lambda,\mu) \, - \, b(\lambda,\mu) \,
                 {\mu (1\!-\!\lambda^2) \over 2(\lambda\!-\!\mu)} \bigg) \;
          \big[ \, C \, , \, L(x,\lambda) \otimes 1 \, \big] \; \delta(x-y)
                                                                      \\[0.1cm]
 & & - \, \bigg( a(\lambda,\mu) \, + \, b(\lambda,\mu) \,
                 {\lambda (1\!-\!\mu^2) \over 2(\lambda\!-\!\mu)} \bigg) \;
          \big[ \, C \,,\, 1 \otimes L(x,\mu) \, \big] \; \delta(x-y) \\[0.1cm]
 & & - \, \bigg( b(\lambda,\mu) \Big( 1 \, + \,
                  {\mu (1\!-\!\lambda^2) \over \lambda\!-\!\mu} \Big) \,
     + \; {2(\lambda\!+\!\mu) \over (1\!-\!\lambda^2)(1\!-\!\mu^2)} \bigg) \;
     \big[ \, j(x) \, , \, L(x,\lambda) \otimes 1 \, \big] \; \delta(x-y)
                                                                      \\[0.1cm]
 & & - \, \bigg( b(\lambda,\mu) \Big( 1 \, - \,
                  {\lambda (1\!-\!\mu^2) \over \lambda\!-\!\mu} \Big) \,
     - \; {2(\lambda\!+\!\mu) \over (1\!-\!\lambda^2)(1\!-\!\mu^2)} \bigg) \;
     \big[ \, j(x) \, , \, 1 \otimes L(x,\mu) \, \big] \; \delta(x-y) \\[0.1cm]
 & & \;\! +~{2(\lambda\!+\!\mu) \over (1\!-\!\lambda^2)(1\!-\!\mu^2)} \,
            \Bigl( j(x) + j(y) \Bigr) \; \delta^\prime(x-y)~~~.
\end{eqnarray*}
Comparing once more, we arrive at the following solution:
\[
 a(\lambda,\mu)~=~- \, {2\lambda\mu \over (1\!-\!\lambda\mu)(\lambda\!-\!\mu)}
                                ~~~,~~~
 b(\lambda,\mu)~=~- \, {2(1 + \lambda\mu)(\lambda - \mu) \over
                        (1 - \lambda\mu)(1 - \lambda^2)(1 - \mu^2)}~~~.
\]
Thus we infer that the Poisson brackets between the $L$-matrices can indeed be
brought into the form (\ref{eq:FPB2}),(\ref{eq:FPB3}) if we choose $r$ and $s$
to be given by
\begin{equation}
 r(z,\lambda,\mu)~
 =~- \, {2\lambda\mu \over (1\!-\!\lambda\mu)(\lambda\!-\!\mu)} \, C \, - \,
        {{2(1\!+\!\lambda\mu)(\lambda\!-\!\mu)} \over
         {(1\!-\!\lambda\mu)(1\!-\!\lambda^2)(1\!-\!\mu^2)}} \, j(z)~~~,
                                                                  \label{eq:r1}
\end{equation}
\begin{equation}
 s(z,\lambda,\mu)~=~- \, {2(\lambda\!+\!\mu) \over
                          (1\!-\!\lambda^2)(1\!-\!\mu^2)} \, j(z)~~~.
                                                                  \label{eq:s1}
\end{equation}
For later use, we also define $d$ to be the difference $s\!-\!r$,
\begin{equation}
 d(z,\lambda,\mu)~=~(s-r)(z,\lambda,\mu)~
 =~{2\lambda\mu \over 1\!-\!\lambda\mu}
   \left( {C \over \lambda\!-\!\mu} \, - \,
          {2 \, j(z) \over \lambda (1\!-\!\mu^2)} \right)~~~,     \label{eq:d1}
\end{equation}
and set
\begin{equation}
 c(z,\lambda,\mu)~=~{2\lambda \over 1\!-\!\lambda^2} \, C~~~.     \label{eq:c1}
\end{equation}
Moreover, the functions $r$, $s$, $d$ and $c$ of one spatial argument are
extended to distributions in two spatial arguments, again denoted by $r$,
$s$, $d$ and $c$, respectively, by multiplying them with a delta function
in the difference of the arguments:
\begin{eqnarray}
 r(x,\lambda\,;y,\mu)~=~r(x,\lambda,\mu) \, \delta(x-y)~&,&~
 s(x,\lambda\,;y,\mu)~=~s(x,\lambda,\mu) \, \delta(x-y)~~~,~~~~~~  \nonumber \\
 d(x,\lambda\,;y,\mu)~=~d(x,\lambda,\mu) \, \delta(x-y)~&,&~
 c(x,\lambda\,;y,\mu)~=~c(x,\lambda,\mu) \, \delta(x-y)~~~.~~~~~~
                                                                \label{eq:rsdc}
\end{eqnarray}

We conclude this section with some remarks on the calculation of Poisson
brackets for transition matrices and monodromy matrices. Our discussion
essentially follows ref.~\cite{Mai1}, to which we refer for further reading.

The Poisson brackets between transition matrices can be obtained from
those between $L$-matrices by making use of a well-known formula for the
variation of the $T$'s induced by a variation of the $L$'s (cf.\ \cite{FaTa},
pp.~191/192)\footnote{In the proof, the case $\, x \!<\! y \,$ can be
reduced to the case $\, x \!>\! y \,$ discussed in \cite{FaTa} by means
of the inversion formula (\ref{eq:TM6}), and of course (\ref{eq:VAR}) is
trivially satisfied if $\, x \!=\! y \,$.}:
\begin{equation}
 \delta T(x,y,\lambda)~=~\int dz~\epsilon(x-y) \, \chi(z\,;x,y) \,
        T(x,z,\lambda) \, \delta L(z,\lambda) \, T(z,y,\lambda)~~~.
                                                                 \label{eq:VAR}
\end{equation}
Here $\epsilon$ is the usual sign function (cf.\ (\ref{eq:EPS})),
while $\, \chi(.\,;x,y) \,$ is the characteristic function of the
interval between $x$ and $y$\footnote{The values $\alpha$ and $\beta$
of $\, \chi(z\,;x,y) \,$ at the points of discontinuity, $z \!=\! x \,$
and $\, z \!=\! y$, are of course irrelevant at this stage: they do
however play a role in the regularization procedure needed to define
Poisson brackets of transition matrices at coinciding points, as will
be explained in more detail below.}:
\begin{equation}
 \chi(z\,;x,y)~~
 =~~\left\{ \begin{array}{ccc}
             \alpha & {\rm for} & z = \min\{x,y\} \\
             1      & {\rm for} & \min\{x,y\} < z < \max\{x,y\} \\
             \beta  & {\rm for} & z = \max\{x,y\} \\
             0      & \multicolumn{2}{c}{\rm otherwise}
            \end{array} \right.~~~.                              \label{eq:CHI}
\end{equation}
Note that
\begin{eqnarray}
 \epsilon(x-y) \; {\partial \over \partial z} \, \chi(z\,;x,y)
 &=& \epsilon(x-y) \left( \delta(z - \min\{x,y\}) \, - \,
                          \delta(z - \max\{x,y\}) \right)~~~~      \nonumber \\
 &=& \delta(z-y) \, - \, \delta(z-x)~~~.                         \label{eq:ECP}
\end{eqnarray}
Now by the chain rule (applied to the functional derivatives involved in the
definition of the Poisson bracket), we have (cf.\ \cite{FaTa}, p.~192)
\begin{eqnarray}
\lefteqn{\{ T(x,y,\lambda) \oktimes T(u,v,\mu) \}}
                                             \hspace{0.3cm} \nonumber \\[0.2cm]
 &=& \int dz \, dw~\epsilon(x-y) \, \epsilon(u-v) \,
     \chi(z\,;x,y) \, \chi(w\,;u,v)                                \nonumber \\
 & & \hphantom{\int dz \, dw~}
     (T(x,z,\lambda) \otimes T(u,w,\mu)) \;
     \{ L(z,\lambda) \oktimes L(w,\mu) \} \;
     (T(z,y,\lambda) \otimes T(w,v,\mu))~.                  \nonumber \\[0.1cm]
 & &                                                            \label{eq:FPB6}
\end{eqnarray}
Similarly,
\begin{eqnarray}
\lefteqn{\{ T(x,y,\lambda) \oktimes L(z,\mu) \}}
                                             \hspace{0.3cm} \nonumber \\[0.2cm]
 &=& \int dz^\prime~\epsilon(x-y) \, \chi(z^\prime;x,y) \;
     (T(x,z^\prime,\lambda) \otimes 1) \;
     \{ L(z^\prime,\lambda) \oktimes L(z,\mu) \} \;
     (T(z^\prime,y,\lambda) \otimes 1)~.                    \nonumber \\[0.1cm]
 & &                                                            \label{eq:FPB7}
\end{eqnarray}
Inserting the Poisson bracket relation (\ref{eq:FPB3}), transforming the
$\delta^\prime$-term into a $\delta$-term by partial integration and using
the differential equations (\ref{eq:TM2}),(\ref{eq:TM3}) together with eqn
(\ref{eq:ECP}), we can perform the integral over $z^\prime$, with the result
\begin{eqnarray}
\lefteqn{\{ T(x,y,\lambda) \oktimes L(z,\mu) \}}
                                             \hspace{0.3cm} \nonumber \\[0.3cm]
 &=& - ~ 2 \left( \delta(z-x) \, - \, \delta(z-y) \right) \,
         (T(x,z,\lambda) \otimes 1) \, s(z,\lambda,\mu) \,
         (T(z,y,\lambda) \otimes 1)                         \nonumber \\[0.2cm]
 & & + ~ \epsilon(x-y) \, \chi(z\,;x,y) \; (T(x,z,\lambda) \otimes 1)
                                                                   \nonumber \\
 & & \phantom{+ ~}
     \Big\{ \, (r+s)^\prime(z,\lambda,\mu) \, - \,
     \big[ \, (r+s)(z,\lambda,\mu) \, , \,
           L(z,\lambda) \otimes 1 \, + \, 1 \otimes L(z,\mu) \, \big] \, \Big\}
                                                                   \nonumber \\
 & & \phantom{+ ~\Big\{ \, (r+s)^\prime(z,\lambda,\mu) \, - \,
                 \big[ \, (r+s)(z,\lambda,\mu) \, , \,
                       L(z,\lambda) \otimes 1 \, + \,}
     (T(z,y,\lambda) \otimes 1)~.~~~~~~                     \nonumber \\[0.1cm]
 & &                                                            \label{eq:FPB8}
\end{eqnarray}
Finally,
\begin{eqnarray}
\lefteqn{\{ T(x,y,\lambda) \oktimes T(u,v,\mu) \}}
                                             \hspace{0.3cm} \nonumber \\[0.2cm]
 &=& \int dz~\epsilon(u-v) \, \chi(z\,;u,v) \;
     (1 \otimes T(u,z,\mu)) \;
     \{ T(x,y,\lambda) \oktimes L(z,\mu) \} \;
     (1 \otimes T(z,v,\mu))~.                               \nonumber \\[0.1cm]
 & &                                                            \label{eq:FPB9}
\end{eqnarray}
Inserting the Poisson bracket relation (\ref{eq:FPB8}) just obtained and using
the differential equations (\ref{eq:TM2}),(\ref{eq:TM3}), we arrive at
\begin{eqnarray}
\lefteqn{\{ T(x,y,\lambda) \oktimes T(u,v,\mu) \}}
                                             \hspace{0.3cm} \nonumber \\[0.2cm]
 &=& - \, 2 \, \epsilon(u-v) \, \chi(z\,;u,v)                      \nonumber \\
 & & \phantom{- \, 2 \,}
     (T(x,z,\lambda) \otimes T(u,z,\mu)) \; s(z,\lambda,\mu) \;
     (T(z,y,\lambda) \otimes T(z,v,\mu)) \; \Big|_{z=y}^{z=x}      \nonumber \\
 & & + \int dz~\epsilon(x-y) \, \epsilon(u-v) \,
       \chi(z\,;x,y) \, \chi(z\,;u,v)                      \nonumber \\[-0.1cm]
 & & \hphantom{+ \int dz~}{\partial \over \partial z}
     \bigg( (T(x,z,\lambda) \otimes T(u,z,\mu)) \;
            (r+s)(z,\lambda,\mu) \;
            (T(z,y,\lambda) \otimes T(z,v,\mu)) \bigg)~,    \nonumber \\[0.1cm]
 & &                                                           \label{eq:FPB10}
\end{eqnarray}
where the integral over $z$ can be performed using eqn (\ref{eq:ECP}): the
result is eqn (\ref{eq:FPB5}).

\goodbreak

The Poisson brackets between monodromy matrices are usually derived by
specializing the Poisson brackets between transition matrices to the case
of coinciding intervals and then sending the boundaries of the interval to
infinity -- a procedure which is known to work well in the ultralocal case.
In general, however, expressions such as
$~\{ L(x,\lambda) \oktimes L(y,\mu) \} \,$,
$~\{ T(x,y,\lambda) \oktimes L(z,\mu) \}~$ and
$~\{ T(x,y,\lambda) \oktimes T(u,v,\mu) \}~$
are distributions in the respective variables, with singular support on the
set of points where (at least) two of their arguments coincide (a set which,
geometrically, is a union of hyperplanes). Therefore, defining Poisson brackets
between transition matrices for coinciding intervals or for adjacent intervals
will normally require some kind of regularization. Now in the ultralocal case,
$\, \{ L(x,\lambda) \oktimes L(y,\mu) \}~$ has only a $\delta$-singularity at
$\, y \!=\! x \,$ and hence\footnote{Roughly speaking, the Poisson brackets
between the $T$'s are obtained from those between the $L$'s by a twofold
integration: this is reflected in the nature of the singularities.}
\addtocounter{footnote}{-1}
$~\{ T(x,y,\lambda) \oktimes T(u,v,\mu) \}~$ is a continuous function
everywhere, so that no regularization is needed. But in the situation
under consideration here, $\, \{ L(x,\lambda) \oktimes L(y,\mu) \}~$ has a
$\delta^\prime$-singularity at $\, y \!=\! x \,$ and hence\footnotemark\
$~\{ T(x,y,\lambda) \oktimes T(u,v,\mu) \}~$ is discontinuous, with a finite
jump proportional to the relevant value of $s$, whenever $\, u \!=\! x \,$
or $\, u \!=\! y \,$ or $\, v \!=\! x \,$ or $\, v \!=\! y$. The simplest
possibility to regularize this expression at the points of discontinuity is
to just take the average over all possible boundary values: this ``total
symmetrization rule'' is the prescription employed by Maillet \cite{Mai1}.
It leads to
\begin{eqnarray}
\lefteqn{\{ T(x,y,\lambda) \oktimes T(x,y,\mu) \}}
                                            \hspace{1.65cm} \nonumber \\[0.1cm]
 &=& \epsilon(x-y) \Big(
     r(x,\lambda,\mu) \; (T(x,y,\lambda) \otimes T(x,y,\mu))
                                                           \nonumber \\[-0.1cm]
 & & \hphantom{\epsilon(xy) \Big( r(x,\lambda,\mu) \; \!} - \,
     (T(x,y,\lambda) \otimes T(x,y,\mu)) \; r(y,\lambda,\mu) \Big)~,~~~~~~
                                                               \label{eq:FPB11}
\end{eqnarray}
and to
\begin{eqnarray}
 \{ T(x,y,\lambda) \oktimes T(y,z,\mu) \}
 &=& \epsilon(x-z) \; (T(x,y,\lambda) \otimes 1) \; s(y,\lambda,\mu) \;
                      (1 \otimes T(y,z,\mu))~.~~~~~~~~         \label{eq:FPB12}
\end{eqnarray}
Obviously, eqn (\ref{eq:FPB11}) implies
\begin{equation}
 \{ T(\lambda) \oktimes T(\mu) \}~~
 =~~r_+(\lambda,\mu) \; (T(\lambda) \otimes T(\mu)) \, - \,
    (T(\lambda) \otimes T(\mu)) \; r_-(\lambda,\mu)            \label{eq:FPB13}
\end{equation}
with
\begin{equation}
 r_\pm(\lambda,\mu)~~=~~\lim_{z \to \pm \infty} r(z,\lambda,\mu)~~~.
\end{equation}

%

\noindent
A slightly more general regularization procedure would be to consider
(\ref{eq:FPB5}) not as an equation between distributions but rather as one
between functions, i.e., to postulate its validity even at the points of
discontinuity. In this way, the (so far irrelevant) parameters $\alpha$
and $\beta$ introduced in eqn (\ref{eq:CHI}) acquire a definite meaning.
They are however not independent, because requiring the derivation rule
\begin{eqnarray}
\lefteqn{\{ T(x,y,\lambda) \, T(y,z,\lambda) \oktimes A \}}
                                             \hspace{0.3cm} \nonumber \\[0.1cm]
 &=& \{ T(x,y,\lambda) \oktimes A \} \; (T(y,z,\lambda) \otimes 1) \; + \;
     (T(x,y,\lambda) \otimes 1) \; \{ T(y,z,\lambda) \oktimes A \}~~~~~~
                                                                \label{eq:DERR}
\end{eqnarray}
\goodbreak \noindent
to be valid, e.g., with $\; A = L(w,\mu) \,$, not only in the sense
of distributions but also in the sense of functions, i.e., even at
$\, w \!=\! y$, forces them to satisfy the constraint \mbox{$~\alpha \!+\!
\beta = 1 \,$.} \linebreak In particular, this constraint holds for the
``total symmetrization rule'' referred to above, which corresponds to the
simplest choice $\, \alpha \!=\! {1\over 2} \,$ and $\, \beta \!=\!
{1\over 2}$, giving
\[
 \chi(z\,;x,y)~~=~~\theta(z - \min\{x,y\})~\theta(\max\{x,y\} - z)
\]
with $\; \theta(0) = {1\over 2} \,$. More general prescriptions will be
discussed in the next section.

The main difficulty with the above derivation of a priori ill-defined Poisson
brackets is the one associated with any regularization procedure: one cannot be
sure that algebraic relations between the unregularized quantities remain valid
for the regularized expressions. This is a well-known and fundamental problem
in
quantum field theory, being the origin, e.g., for the occurrence of anomalies.
In the present case, the algebraic relation in question is the Jacobi identity.
In particular, we have seen that the monodromy matrices are subject to the
standard Poisson bracket relations (\ref{eq:FPB13}) of the $r$-matrix approach
\cite{FaTa}, but with asymptotic $r$-matrices $r_+$ and $r_-$ which --
according
to the classification theorem given in the appendix -- do {\em not} satisfy the
classical Yang-Baxter equation, i.e., the relation normally imposed in order to
guarantee that the Jacobi identity holds. This apparent paradox can be resolved
by remembering that the lhs.\ of eqn (\ref{eq:FPB13}) is a regularized Poisson
bracket and noting that of course the corresponding Jacobi identity must be
regularized as well. In fact, the regularized Jacobi identity will involve
regularized double Poisson brackets which are not identical with the double
Poisson brackets one would normally obtain from a naive twofold application
of eqn (\ref{eq:FPB13}), and hence its validity will be governed by an equation
which is {\em not} identical with the classical Yang-Baxter equation. To derive
it, one must go back to the definitions and must compute double Poisson
brackets
between transition matrices, which necessarily involve simple Poisson brackets
between transition matrices and the matrices $\; r \pm s \,$: their calculation
will be one of our goals in the next section.

Before proceeding, we would like to point out that an alternative and more
direct method for deriving Poisson brackets between transition matrices at
coinciding points, without recourse to an explicit regularization, has been
proposed in the recent literature \cite{FrMa}. It is based on expressing
the $L$-matrix as a gauge transform of some other matrix in such a way that
both this other matrix and the gauge transformation itself satisfy Poisson
bracket relations of ultralocal type. This approach does work for the principal
chiral models, in which the coefficient of the classical Schwinger term is both
field independent and central -- at least as long as one considers only left
currents or only right currents. But it is not clear at all whether the
technique can be extended to the situation of interest here, which is
substantially more general. This is certainly an interesting question
but also quite a difficult one, which warrants a separate investigation.
For the time being, we have preferred to stick to the more traditional method.

\section{The Lie-Poisson Structure}

In the preceding section, we have rewritten the fundamental Poisson brackets
between $L$-matrices in tensor notation and have discussed some consequences
that can be drawn, such as the calculation of the fundamental Poisson brackets
between transition matrices and monodromy matrices. Our goal now is to exhibit
the full Lie-Poisson structure of the theory -- which in terms of components
is contained in eqns (\ref{eq:PB1a}) and (\ref{eq:PB2a}) (together with eqn
(\ref{eq:CA6})) -- and to discuss the algebraic constraints resulting from
the Jacobi identity for Poisson brackets. This requires, first of all, a
slight modification of the tensor notation used previously, adapted to deal
not only with the tensor product $\, U(\gotg\,) \otimes U(\gotg\,) \,$ of
$U(\gotg\,)$ with itself but with its tensor powers $U(\gotg\,)^{\otimes n}$
of arbitrary order and with the various ways in which these can be embedded
into each other. For example, note that $U(\gotg\,)$ and $\, U(\gotg\,) \otimes
U(\gotg\,) \,$ can be embedded into $U(\gotg\,)^{\otimes n}$ according to
\[
U(\gotg\,) \longrightarrow U(\gotg\,)^{\otimes n}~~,~~u \longmapsto u_k
{}~=~1 \smotimes \ldots \smotimes u \smotimes \ldots \smotimes 1
\]
(with $u$ appearing in the $k^{\rm th}$ place, $1 \leq k \leq n$) and to
\[
U(\gotg\,) \otimes U(\gotg\,) \longrightarrow U(\gotg\,)^{\otimes n}~~,~~
u \smotimes v \longmapsto (u \smotimes v)_{kl}
{}~=~1 \smotimes \!\ldots\! \smotimes u \smotimes \!\ldots\! \smotimes
                                    v \smotimes \!\ldots\! \smotimes 1
\]
(with $u$ appearing in the $k^{\rm th}$ place and $v$ appearing in the
$l^{\rm th}$ place, $1 \leq k,l \leq n$), respectively. Moreover, we shall
find it convenient to introduce, besides the composite field $j$ used so far,
a new composite field $\sigma$, the two being related by $\mbox{$~\sigma
= 1 - 2j~$}$ when both are considered as taking values in the space
$~L(\mbox{\goth g}\,) \cong \mbox{\goth g} \, \otimes \mbox{\goth g}^\ast~$
of linear transformations on $\mbox{\goth g}\,$ (with $1$ denoting the identity
on $\mbox{\goth g}\,$), or equivalently, by $\mbox{$~\sigma = C - 2j~$}$ when
both are considered as taking values in $\, U(\gotg\,) \otimes U(\gotg\,) \,$.
Then for $\, k,l,m \,$ all distinct, we have
\begin{eqnarray*}
&[ \, C_{kl} \, , \, C_{km} \, ]~~
 =~~f^{abc} \; (T_a)_k \, (T_b)_l \, (T_c)_m~~~,
                                                                     &\\[0.2cm]
&[ \, C_{kl} \, , \, \sigma_{km} \, ]~~
 =~~f^{ab}_{\phantom{ab} d} \, \sigma^{dc} \; (T_a)_k \, (T_b)_l \, (T_c)_m~~~,
                                                                     &\\[0.2cm]
&[ \, \sigma_{kl} \, , \, \sigma_{km} \, ]~~
 =~~f^a_{de} \, \sigma^{db} \, \sigma^{ec} \; (T_a)_k \, (T_b)_l \, (T_c)_m~~
 =~~f^{bc}_{\phantom{bc} d} \, \sigma^{da} \; (T_a)_k \, (T_b)_l \, (T_c)_m~~~,
                                                                            &
\end{eqnarray*}
where the second equality in the last equation is based on the fact that we
are dealing with a symmetric space, so that $\sigma$ is a (field dependent)
involutive automorphism of $\mbox{\goth g}\,$. In particular, we have, for
$\, k,l,m,n \,$ all distinct, the following algebraic identities:
\begin{eqnarray}
 [ \, C_{kl} \, , \, C_{km} \, ]
 &=&  - \, [ \, C_{kl} \, , \, C_{lm} \, ]~~~,         \label{eq:ID7} \\[0.2cm]
 [ \, C_{kl} \, , \, \sigma_{km} \, ]
 &=&  - \, [ \, C_{kl} \, , \, \sigma_{lm} \, ]~~~,    \label{eq:ID8} \\[0.2cm]
 [ \, \sigma_{kl} \, , \, \sigma_{km} \, ]
 &=&  - \, [ \, \sigma_{kl} \, , \, C_{lm} \, ]~~~,              \label{eq:ID9}
\end{eqnarray}
\begin{equation}
  [ \, C_{kl} \, , \, C_{mn} \, ]~=~
  [ \, C_{kl} \, , \, \sigma_{mn} \, ]~=~
  [ \, \sigma_{kl} \, , \, \sigma_{mn} \, ]~=~0~~~.             \label{eq:ID10}
\end{equation}

In this notation, the Poisson brackets for the $L$'s and $j$'s given by eqns
(\ref{eq:PB1a})--(\ref{eq:PB3a}) read
\begin{eqnarray}
 \{ L_k(x_k,\lambda_k) \, , \, L_l(x_l,\lambda_l) \}
 &\!=\!& \left[ \, {2 \, C_{kl} \over \lambda_k\!-\!\lambda_l} \; , \;
         {\lambda_l^2 \over 1\!-\!\lambda_l^2} \, L_k(x_k,\lambda_k) \, + \,
         {\lambda_k^2 \over 1\!-\!\lambda_k^2} \, L_l(x_l,\lambda_l) \,
         \right] \, \delta(x_k-x_l)                         \nonumber \\[0.1cm]
 &\!   & + \; {4 \over (1\!-\!\lambda_k^2)(1\!-\!\lambda_l^2)} \,
         \Bigl( \lambda_l j_{kl}(x_k) + \lambda_k j_{kl}(x_l)
         \Bigr) \, \delta^\prime(x_k-x_l)~,~~~~~~     \label{eq:PB1f} \\[0.2cm]
 \{ L_k(x_k,\lambda_k) \, , \, j_{lm}(x_l) \}
 &\!=\!& {2\lambda_k \over 1\!-\!\lambda_k^2} \,
         \Bigl[ \, C_{kl} + C_{km} \, , \, j_{lm}(x_k) \, \Bigr] \;
         \delta(x_k-x_l)~,~~~~~~                      \label{eq:PB2e} \\[0.2cm]
 \{ j_{kl}(x_k) \, , \, j_{mn}(x_m) \} &\!=\!& 0~.~~~~~~        \label{eq:PB3e}
\end{eqnarray}
Observe that due to the identity (\ref{eq:ID7}), eqn (\ref{eq:PB2e}) continues
to hold when $j$ is replaced by $\sigma$ or by any other linear combination
of $C$ and $j$, such as $r$, $s$ or $d$. Therefore, the Poisson brackets
for the $D$'s and $d$'s (cf.\ (\ref{eq:LO}), (\ref{eq:FPB4}),
(\ref{eq:d1})--(\ref{eq:rsdc})) take the following, much more transparent form:
\setlength{\arraycolsep}{0pt}
\begin{eqnarray}
 & &\{ D_k(x_k,\lambda_k) \, , \, D_l(x_l,\lambda_l) \}     \nonumber \\[0.1cm]
 & &\hspace{2.1cm} =~\;\left[ \, d_{kl}(x_k,\lambda_k\,;x_l,\lambda_l) \, , \,
                                 D_k(x_k,\lambda_k) \, \right] \, - \,
                       \left[ \, d_{lk}(x_l,\lambda_l\,;x_k,\lambda_k) \, , \,
                                 D_l(x_l,\lambda_l) \,
                       \right]~,~~~~~~                \label{eq:PB1g} \\[0.2cm]
 & &\{ D_k(x_k,\lambda_k) \, , \, d_{lm}(x_l,\lambda_l\,;x_m,\lambda_m) \}
                                                            \nonumber \\[0.1cm]
 & &\hspace{2.1cm} =~\;\left[ \, c_{kl}(x_k,\lambda_k\,;x_l,\lambda_l) +
                                 c_{km}(x_k,\lambda_k\,;x_m,\lambda_m) \, , \,
                                 d_{lm}(x_l,\lambda_l\,;x_m,\lambda_m) \,
                       \right]~,~~~~~~                \label{eq:PB2g} \\[0.2cm]
 & &\{ d_{kl}(x_k,\lambda_k\,;x_l,\lambda_l) \, , \,
       d_{mn}(x_m,\lambda_m\,;x_n,\lambda_n) \}~~=~~0~,         \label{eq:PB3g}
\end{eqnarray}
where, as elements of $U(\gotg\,)^{\otimes n}$, $\, d_{lk} = d_{kl} \,$ and
$~c_{lk} = c_{kl}~$, because the tensors $C$ and $j$ are symmetric. Obviously,
these formulas can be further simplified by letting the indices $k,l,m$ refer
not only to the position inside the tensor product but also to the spatial
variable and to the spectral parameter, so $k,l$ and $m$ now stand for
$(k,x_k,\lambda_k)$, $(l,x_l,\lambda_l)$ and $(m,x_m,\lambda_m)$,
respectively. In these composite indices, $d$ and $c$ are of course no
longer symmetric, $\, d_{lk} \neq d_{kl} \,$, $c_{lk} \neq c_{kl} \,$,
but (\ref{eq:PB1g})--(\ref{eq:PB3g}) reduce to
\setlength{\arraycolsep}{5pt}
\begin{eqnarray}
 \{ D_k \, , \, D_l \}
 &=& \left[ \, d_{kl} \, , \, D_k \, \right] \, - \,
     \left[ \, d_{lk} \, , \, D_l \, \right]~~~,      \label{eq:PB1h} \\[0.2cm]
 \{ D_k \, , \, d_{lm} \}
 &=& \left[ \, c_{kl} + c_{km} \, , \, d_{lm} \, \right]~~~,
                                                      \label{eq:PB2h} \\[0.2cm]
 \{ d_{kl} \, , \, d_{mn} \} &=& 0~~~.                          \label{eq:PB3h}
\end{eqnarray}
The first of these equations has appeared in the literature before \cite{BaVi},
but the others seem to be new. Together, they show that the $D$'s and $d$'s
generate an algebra which closes under Poisson brackets, because $c$ is a
numerical (i.e., field independent) matrix (cf.\ (\ref{eq:c1})). In the theory
of non-ultralocal integrable models of the type considered here, this algebra
plays a central role: it is the analogue of the classical Yang-Baxter algebra
which is relevant to ultralocal integrable models such as, e.g., the non-linear
Schr\"odinger equation \cite{FaTa}. But in order to give a mathematically more
precise interpretation, one must make a clear-cut distinction between the
abstract algebra and its concrete representations. For the classical
Yang-Baxter algebras, this distinction is well understood: the structure
of the abstract algebra is reflected in the $r$-matrix, which satisfies the
classical Yang-Baxter equation (\ref{eq:YBE1}), while the $L$'s define a
concrete representation of that algebra by functionals on the phase space
of the theory, according to eqn (\ref{eq:FPB1}). Here, the structure of the
abstract algebra is (at least partially) reflected in the $c$-matrix,
while the $D$'s and $d$'s define a concrete representation of that
algebra by functionals on the phase space of the theory, according to
eqns (\ref{eq:PB1g})--(\ref{eq:PB3g}) or (\ref{eq:PB1h})--(\ref{eq:PB3h}).
In particular, the $r$-matrix -- which in this context is the antisymmetric
part of the (field dependent) $d$-matrix -- now has a different meaning.

The investigation of the mathematical structures that underly this new algebra
is still a completely open subject. The first step would be to identify its
defining relations, i.e., the analogue of the Jacobi identity for Lie algebras
or the classical Yang-Baxter equation for classical Yang-Baxter algebras: they
should include a structure equation for the $c$-matrix which we suspect, once
again, to be quadratic. Another problem would be to develop a representation
theory for such algebras, and in particular, study their representations in
Poisson algebras of function(al)s on phase spaces. All of this must be left
to future work. Here, we just want to analyze the consequences of the fact
that eqns (\ref{eq:PB1g})--(\ref{eq:PB3g}) or (\ref{eq:PB1h})--(\ref{eq:PB3h})
must be consistent with the Jacobi identity for Poisson brackets.

In order to do so, we first rewrite the definition (\ref{eq:d1}) of $d$
in terms of $C$ and $\sigma$, rather than $C$ and $j$:
\begin{equation}
 d(z,\lambda,\mu)~=~(s-r)(z,\lambda,\mu)~
 =~{2\mu \over 1\!-\!\mu^2}
   \left( {\mu C \over \lambda\!-\!\mu} \, + \,
          {\sigma(z) \over 1\!-\!\lambda\mu} \right)~~~.          \label{eq:d2}
\end{equation}
Moreover, we shall find it convenient to use the following general notation.
Given an arbitrary associative algebra with unit ${\cal A}\,$, we extend the
tensor notation introduced above from $U(\gotg\,)$ to ${\cal A}\,$ and define,
for $\; t \smin {\cal A} \otimes {\cal A} \;$ and $\, k,l,m \,$ all distinct,
\begin{equation}
 {\rm YB}(t)_{klm}~~=~~[ \, t_{kl} \, , \, t_{km} \, ] \; + \;
                       [ \, t_{kl} \, , \, t_{lm} \, ] \; - \;
                       [ \, t_{km} \, , \, t_{ml} \, ]~~~.       \label{eq:DYB}
\end{equation}
Note the somewhat unusual position of the indices in the last term, which has
however appeared in the literature before \cite{AvTa,BaVi,Mai1}. Of course,
for antisymmetric $t$ (and with $\; \{ k,l,m \} = \{ 1,2,3 \}$), this
expression reduces to the familiar lhs.\ of the classical Yang-Baxter
equation
\begin{equation}
 {\rm YB}(t)_{klm}~~=~~0~~~.                                    \label{eq:YBE2}
\end{equation}
In Appendix \ref{AppendixA}, we compute ${\rm YB}(t)$ for $t$ an arbitrary
linear combination of $C$ and $\sigma$, with coefficients depending on the
spectral parameters, and give a classification of all solutions to the
classical Yang-Baxter equation which are of this type (Theorem \ref{th:A1}).
The result has already been obtained in ref.~\cite{AvTa}: essentially, we
slightly extend the proof given there in order to allow for solutions with
singularities.

Turning to the verification of the Jacobi identity, we first of all use
the derivation rule for the Poisson bracket $\, \{.\,,.\} \,$ and the
Jacobi identity for the commutator $\, [.\,,.] \,$ to derive, from eqn
(\ref{eq:PB1h}), the following relation
\cite{BaVi,Mai1}:
\begin{eqnarray}
\lefteqn{\{ \, D_k \, , \, \{ \, D_l \, , \, D_m \, \} \, \}~+~{\rm cyclic}}
                                                    \hspace{0.5cm} \nonumber \\
 &=& [ \; D_k \; , \; {\rm YB}(d)_{klm} \, + \,
                      \{ \, D_l \, , \, d_{km} \, \} \, - \,
                      \{ \, D_m \, , \, d_{kl} \, \} \; ]
     ~+~{\rm cyclic}~.~~~~~~                                    \label{eq:DDD1}
\end{eqnarray}
Inserting eqn (\ref{eq:PB2h}), this becomes
\begin{eqnarray}
\lefteqn{\{ \, D_k \, , \, \{ \, D_l \, , \, D_m \, \} \, \}~+~{\rm cyclic}}
                                                    \hspace{0.5cm} \nonumber \\
 &=& [ \; D_k \; , \; {\rm YB}(d)_{klm} \, + \,
                      [ \, c_{lk} + c_{lm} \, , \, d_{km} \, ] \, - \,
                      [ \, c_{mk} + c_{ml} \, , \, d_{kl} \, ] \; ]
     ~+~{\rm cyclic}~.~~~~~~                                    \label{eq:DDD2}
\end{eqnarray}
Thus the rhs.\ of this expression must vanish in order for the Jacobi identity
to be satisfied. But actually, more than this is true: namely, we have
\begin{equation}
 {\rm YB}(d)_{klm} \, + \, \{ \, D_l \, , \, d_{km} \, \} \, - \,
                           \{ \, D_m \, , \, d_{kl} \, \}~~=~~0~~~,
                                                                \label{eq:YBE3}
\end{equation}
or equivalently,
\begin{equation}
 {\rm YB}(d)_{klm} \, + \, [ \, c_{lk} + c_{lm} \, , \, d_{km} \, ] \, - \,
                           [ \, c_{mk} + c_{ml} \, , \, d_{kl} \, ]~~=~~0~~~,
                                                                \label{eq:YBE4}
\end{equation}
as can be checked by an explicit calculation using the definitions
(\ref{eq:c1}) and (\ref{eq:d2}) of $c$ and $d$, together with the
identities (\ref{eq:ID7})--(\ref{eq:ID9}) (cf.\ eqn (\ref{eq:YBE7})
in Appendix \ref{AppendixA}): this confirms, for the class of models
under consideration here, the validity of the ``extended dynamical
Yang-Baxter relation'' postulated by Maillet \cite{Mai1}. (Indeed,
if we write, as in eqn (2.19) there,
\begin{eqnarray*}
\lefteqn{\{ \, D_k(x_k,\lambda_k) \, , \,
               d_{lm}(x_l,\lambda_l\,;x_m,\lambda_m) \, \}}   \hspace{1.8cm} \\
 &=& \{ \, L_k(x_k,\lambda_k) \, , \,
           (s+r)_{ml}(x_m;\lambda_m,\lambda_l) \, \} \, \delta(x_l-x_m)      \\
 &=& - \, H_{k,ml}(x_k;\lambda_k,\lambda_m,\lambda_l) \;
          \delta(x_k-x_m) \, \delta(x_l-x_m)~~~,
\end{eqnarray*}
and recall that our $L$ corresponds to $-L$ there, we see that
our equation (\ref{eq:YBE3}), after splitting off a factor
$~\delta(x_k-x_l) \, \delta(x_k-x_m) \,$, becomes identical
with eqn (2.18) there.) \\
Similarly, using once again the derivation rule for the Poisson
bracket $\, \{.\,,.\} \,$ and the Jacobi identity for the commutator
$\, [.\,,.] \,$, we obtain from eqns (\ref{eq:PB2h})--(\ref{eq:PB3h})
\begin{eqnarray}
\lefteqn{\{ \, D_k \, , \, \{ \, D_l \, , \, d_{mn} \, \} \, \}~+~
         \{ \, D_l \, , \, \{ \, d_{mn} \, , \, D_k \, \} \, \}~+~
         \{ \, d_{mn} \, , \, \{ \, D_k \, , \, D_l \, \} \, \}}
                                                    \hspace{0.5cm} \nonumber \\
 &=& [ \; d_{mn} \; , \;
     [ \, c_{km} + c_{kn} \, , \, c_{lm} + c_{ln} \, ]~-~
     [ \, c_{km} + c_{kn} \, , \, d_{kl} \, ]~+~
     [ \, c_{lm} + c_{ln} \, , \, d_{lk} \, ] \; ]~.~~~~~~~~     \label{eq:DDd}
\end{eqnarray}
Again the rhs.\ of this expression must vanish in order for the Jacobi identity
to be satisfied, and again this can be checked by an explicit calculation using
the definitions (\ref{eq:c1}) and (\ref{eq:d2}) of $c$ and $d$, together with
the identities (\ref{eq:ID7})--(\ref{eq:ID10}) and
\begin{eqnarray}
 {[}\, [ \, C_{km} + C_{kn} \, , \, \sigma_{kl} \, ] \, , \, C_{mn} \, ]~~
 =~~{[} \, [ \, C_{lm} + C_{ln} \, , \, \sigma_{kl} \, ] \, , \, C_{mn} \, ]
 &=& 0~~~,                                                   \label{eq:ID11} \\
 {[}\, [ \, C_{km} - C_{ln} \, , \, \sigma_{kl} \, ] \, , \, \sigma_{mn} \, ]~~
 =~~{[}\, [ \, C_{kn} - C_{lm} \, , \, \sigma_{kl} \, ] \, , \, \sigma_{mn} \,
]
 &=& 0~~~,                                                      \label{eq:ID12}
\end{eqnarray}
the second of which holds because
\begin{equation}
 {[}\, [ \, C_{km} \, , \, \sigma_{kl} \, ] \, , \, \sigma_{mn} \, ]~~
 =~~- \, [ \, [ \, \sigma_{kn} \,,\, \sigma_{mn} \, ] \,,\, \sigma_{kl} \, ]~~
 =~~[ \, [ \, C_{ln} \, , \, \sigma_{kl} \, ] \, , \, \sigma_{mn} \, ]~.~~~~~~
                                                                \label{eq:ID13}
\end{equation}
Finally, in the remaining combinations
\[
  \{ \, D_i \, , \, \{ \, d_{kl} \, , \, d_{mn} \, \} \, \}~+~
  \{ \, d_{kl} \, , \, \{ \, d_{mn} \, , \, D_i \, \} \, \}~+~
  \{ \, d_{mn} \, , \, \{ \, D_i \, , \, d_{kl} \, \} \, \}
\vspace{-0.2cm}
\]
and
\[
  \{ \, d_{ij} \, , \, \{ \, d_{kl} \, , \, d_{mn} \, \} \, \}~+~
  \{ \, d_{kl} \, , \, \{ \, d_{mn} \, , \, d_{ij} \, \} \, \}~+~
  \{ \, d_{mn} \, , \, \{ \, d_{ij} \, , \, d_{kl} \, \} \, \}
\vspace{0.2cm}
\]
each term vanishes separately.

Using the $c$'s and $d$'s, we can write down a closed algebra not only for
the Lax operators $D(x,\lambda)$ but also for the transition matrices
$\, T(x,y,\lambda) \,$ -- at least when the relations involved are viewed
as equations between distributions in all spatial variables that appear,
so that no regularization is required. To this end, we continue to use the
tensor notation introduced above, which -- in view of the fact that the
transition matrices take values in the Lie group $G$ -- we can do provided
we replace the universal enveloping algebra $U(\gotg\,)$ by a `complete'
algebra $\hat{U}(\gotg\,)$ which incorporates both $U(\gotg\,)$ and $G$:
this can be achieved by representing $U(\gotg\,)$ as the algebra of left
invariant differential operators \cite[p.~108]{Hel} and $G$ as the group
of left translation operators on the space $C^\infty(G)$ of real-valued
smooth functions on $G$: $\hat{U}(\gotg\,)$ can then be defined as the
subalgebra of linear operators on $C^\infty(G)$ generated by $U(\gotg\,)$
and $G$. With this notation, the Poisson bracket relation (\ref{eq:PB2e})
between the $L$'s and the $j$'s gives rise to a Poisson bracket relation
between the $T$'s and the $j$'s, according to
\begin{eqnarray}
\lefteqn{\{ T_k(x,y,\lambda) \, , \, j_{lm}(z) \}}
                                             \hspace{0.5cm} \nonumber \\[0.2cm]
 &=& \int dz^\prime~\epsilon(x-y) \, \chi(z^\prime;x,y) \;
     T_k(x,z^\prime,\lambda) \;
     \{ L_k(z^\prime,\lambda) \, , \, j_{lm}(z) \} \;
     T_k(z^\prime,y,\lambda)~,                              \nonumber \\[0.1cm]
 & &                                                           \label{eq:FPB15}
\end{eqnarray}
(cf.\ eqns (\ref{eq:FPB6}),(\ref{eq:FPB7}),(\ref{eq:FPB9})), namely
\begin{eqnarray}
\lefteqn{\{ T_k(x,y,\lambda) \, , \, j_{lm}(z) \}}
                                             \hspace{0.5cm} \nonumber \\[0.2cm]
 &=& {2\lambda \over 1\!-\!\lambda^2} \;
     \epsilon(x-y) \, \chi(z\,;x,y) \; T_k(x,z,\lambda) \;
     [ \, C_{kl} + C_{km} \, , \, j_{lm}(z) \, ] \; T_k(z,y,\lambda)~.
                                                            \nonumber \\[0.1cm]
 & &                                                           \label{eq:FPB16}
\end{eqnarray}
Observe again that due to the identity (\ref{eq:ID7}), this equation continues
to hold when $j$ is replaced by $\sigma$ or by any other linear combination of
$C$ and $j$, such as $r$, $s$ or $d$. Therefore, the basic Poisson brackets
for the $T$'s and $d$'s (cf.\ (\ref{eq:FPB5}),(\ref{eq:d1}),(\ref{eq:c1}))
take the following form:
\goodbreak
\begin{eqnarray}
\lefteqn{\{ T_k(x_k,y_k,\lambda_k) \, , \, T_l(x_l,y_l,\lambda_l) \}}
                                             \hspace{0.5cm} \nonumber \\[0.2cm]
 &=& +~\epsilon(x_k-y_k) \, \chi_{kl}(z\,;x_k,y_k)                 \nonumber \\
 & & \phantom{+~}
     T_k(x_k,z,\lambda_k) \, T_l(x_l,z,\lambda_l) \;
     d_{lk}(z,\lambda_l,\lambda_k) \;
     T_k(z,y_k,\lambda_k) \, T_l(z,y_l,\lambda_l) \;
     \Big|_{z=y_l}^{z=x_l}                                  \nonumber \\[0.1cm]
 & & -~\epsilon(x_l-y_l) \, \chi_{lk}(z\,;x_l,y_l)                 \nonumber \\
 & & \phantom{+~}
     T_k(x_k,z,\lambda_k) \, T_l(x_l,z,\lambda_l) \;
     d_{kl}(z,\lambda_k,\lambda_l) \;
     T_k(z,y_k,\lambda_k) \, T_l(z,y_l,\lambda_l) \;
     \Big|_{z=y_k}^{z=x_k}~,~~~~~~                    \label{eq:PB4a} \\[0.3cm]
\lefteqn{\{ T_k(x,y,\lambda_k) \, , \, d_{lm}(z,\lambda_l,\lambda_m) \}}
                                             \hspace{0.5cm} \nonumber \\[0.2cm]
 &=& \epsilon(x-y) \, \chi_{klm}(z\,;x,y)                   \nonumber \\[0.1cm]
 & & T_k(x,z,\lambda_k) \;
     [ \, c_{kl}(z,\lambda_k,\lambda_l) + c_{km}(z,\lambda_k,\lambda_m) \, , \,
          d_{lm}(z,\lambda_l,\lambda_m) \, ] \;
     T_k(z,y,\lambda_k)~.~~~~~~                                 \label{eq:PB5a}
\end{eqnarray}
As before, the $\chi$'s are characteristic functions of the respective
intervals, their values at the points of discontinuity being allowed to
depend on the spectral parameters involved (but not on the field $\varphi$):
\begin{equation}
 \chi_{kl}(z\,;x,y)~~
 =~~\left\{ \begin{array}{ccc}
             \alpha_{kl} \equiv \alpha(\lambda_k,\lambda_l)
                         & {\rm for} & z = \min\{x,y\} \\
             1           & {\rm for} & \min\{x,y\} < z < \max\{x,y\} \\
             \beta_{kl}  \equiv  \beta(\lambda_k,\lambda_l)
                         & {\rm for} & z = \max\{x,y\} \\
             0           & \multicolumn{2}{c}{\rm otherwise}
            \end{array} \right.~,~~                             \label{eq:CHI1}
\end{equation}
\begin{equation}
 \chi_{klm}(z\,;x,y)~~
 =~~\left\{ \begin{array}{ccc}
             \alpha_{klm} \equiv \alpha(\lambda_k,\lambda_l,\lambda_m)
                          & {\rm for} & z = \min\{x,y\} \\
             1            & {\rm for} & \min\{x,y\} < z < \max\{x,y\} \\
             \beta_{klm}  \equiv  \beta(\lambda_k,\lambda_l,\lambda_m)
                          & {\rm for} & z = \max\{x,y\} \\
             0            & \multicolumn{2}{c}{\rm otherwise}
            \end{array} \right.~.~~                             \label{eq:CHI2}
\end{equation}
As mentioned in the previous section, the values of the $\alpha$'s and
$\beta$'s are irrelevant as long as (\ref{eq:PB4a}) and (\ref{eq:PB5a})
are interpreted as equations between distributions. Moreover, we can
use the same reasoning as in ref.~\cite{Mai1} to conclude from eqns
(\ref{eq:PB4a}) and (\ref{eq:PB5a}) that the Jacobi identity holds,
in the sense of distributions, for the algebra generated by the $T$'s
and $d$'s: in particular, it is valid for multiple Poisson brackets
\mbox{$\{ T_k(x_k,y_k,\lambda_k) \, , \{ T_l(x_l,y_l,\lambda_l) \, ,
\ldots \} \} \,$.}

On the other hand, we recall from the previous section that fixing the values
of the $\alpha$'s and $\beta$'s in eqns (\ref{eq:CHI1}),(\ref{eq:CHI2}) amounts
to a specific choice of regularization for the Poisson brackets in the equal
points limit ($\, x_l \! = \! x_k \,$ or $\, x_l \! = \! y_k \,$ or
$\, y_l \! = \! x_k \,$ or $\, y_l \! = \! y_k \,$ in (\ref{eq:PB4a}),
$\, z \! = \! x \,$ or $\, z \! = \! y \,$ in (\ref{eq:PB5a})), which
is required to define Poisson brackets between transition matrices for
coinciding or adjacent intervals, and in particular between monodromy
matrices. Their choice is constrained by the derivation rule (\ref{eq:DERR}),
which leads to
\begin{equation}
 \alpha_{kl} + \beta_{kl}~=~1~~~,~~~\alpha_{klm} + \beta_{klm}~=~1~~~.
                                                                  \label{eq:AB}
\end{equation}
Now since the regularized basic Poisson brackets between $T$'s and $d$'s are
given by products of $T$'s and $d$'s, we are able to compute their multiple
Poisson brackets from repeated application of the derivation rule and the
regularized basic Poisson brackets. We shall refer to such a procedure as
a {\em one-step regularization} -- in contrast, e.g., to Maillet's ``total
symmetrization rule'' \cite{Mai1} which is a {\em multi-step regularization}
insofar as each multiple Poisson bracket requires a separate regularization
that cannot be reduced to the regularization of multiple Poisson brackets
with a smaller number of factors.

In order to see whether the above one-step regularization works, we compute
the lhs.\ of the Jacobi identity for three transition matrices on coinciding
intervals. To begin with, we write down the regularized Poisson bracket of
two transition matrices on coinciding intervals, assuming for simplicity
that $x\!>\!y$:
\begin{eqnarray}
\lefteqn{\{ T_k(x,y,\lambda_k) \, , \, T_l(x,y,\lambda_l) \}}
                                             \hspace{0.4cm} \nonumber \\[0.1cm]
 &=& b_{kl}(x,\lambda_k,\lambda_l) \,
     T_k(x,y,\lambda_k) \, T_l(x,y,\lambda_l) \; - \;
     T_k(x,y,\lambda_k) \, T_l(x,y,\lambda_l) \,
     a_{kl}(y,\lambda_k,\lambda_l)~,~~~~~~                  \nonumber \\[0.1cm]
 & &                                                            \label{eq:PB4b}
\end{eqnarray}
with
\begin{eqnarray}
 a_{kl}(y,\lambda_k,\lambda_l)
 &=& (\alpha_{kl} + \alpha_{lk}) \, r_{kl}(y,\lambda_k,\lambda_l) \; + \;
     (\alpha_{kl} - \alpha_{lk}) \, s_{kl}(y,\lambda_k,\lambda_l)~,~~~~~~
                                                            \nonumber \\[0.1cm]
 b_{kl}(x,\lambda_k,\lambda_l)
 &=& (\beta_{kl} + \beta_{lk}) \, r_{kl}(x,\lambda_k,\lambda_l) \; + \;
     (\beta_{kl} - \beta_{lk}) \, s_{kl}(x,\lambda_k,\lambda_l)~.~~~~~~
                                                                \label{eq:PB4c}
\end{eqnarray}
(Note that in the special case $~\alpha_{kl} = \beta_{kl} = {1\over 2} \,$,
which corresponds to Maillet's ``total symmetrization rule'', eqn
(\ref{eq:PB4b}
with eqn (\ref{eq:PB4c}) reduces to eqn (\ref{eq:FPB11}).)
Now straightforward calculation gives
\begin{eqnarray}
\lefteqn{\{ T_k(x,y,\lambda_k) \, , \,
         \{ T_l(x,y,\lambda_l) \, , \, T_m(x,y,\lambda_m) \}
\}~+~\mbox{cyclic}}
                                             \hspace{0.4cm} \nonumber \\[0.3cm]
 &=& +~T_k(x,y,\lambda_k) \, T_l(x,y,\lambda_l) \, T_m(x,y,\lambda_m)
                                                                   \nonumber \\
 & & \times~\left\{ {\rm YB}(a(y))_{klm} \; +
             \left( {2\lambda_k \over 1\!-\!\lambda_k^2} \;
              \Big[ \, C_{kl} + C_{km} \, , \, \alpha_{klm} \,
                       a_{lm}(y,\lambda_l,\lambda_m) \, \Big] \;
             + \; \mbox{cyclic} \right) \right\}                   \nonumber \\
 & & -~\left\{ {\rm YB}(b(x))_{klm} \; -
        \left( {2\lambda_k \over 1\!-\!\lambda_k^2} \;
         \Big[ \, C_{kl} + C_{km} \, , \, \beta_{klm} \,
                  b_{lm}(x,\lambda_l,\lambda_m) \, \Big] \;
        + \; \mbox{cyclic} \right) \right\}                        \nonumber \\
 & & \times~T_k(x,y,\lambda_k) \, T_l(x,y,\lambda_l) \, T_m(x,y,\lambda_m)~~~.
                                                                  \label{eq:JI}
\end{eqnarray}
It is hard to see how the rhs.\ of this equation can be zero, identically in
$x$, $y$, $\varphi$ and $\dot{\varphi}$, unless the terms in curly brackets
vanish separately. This, however, is excluded by the following no-go theorem:

\begin{theorem} 
 With the above notation, there is {\sl no} choice for the functions
 $\; \alpha_{kl} \, , \, \alpha_{klm} \;$ and $\; \beta_{kl} \, , \,
 \beta_{klm} \,$, subject to the condition (\ref{eq:AB}), such that
 the two terms in curly brackets of eqn (\ref{eq:JI}) vanish separately.
\end{theorem}

\noindent
A proof of this theorem will be given in the appendix.

\section{Appendix} \label{AppendixA}

Our purpose in this appendix is to formulate and prove the classification
theorem mentioned in Sect.~4 and to present a proof of the no-go-theorem
given at the end of Sect.~4.

\subsection{Classification Theorem}

The classification theorem concerns solutions $t$ of the classical Yang-Baxter
equation\footnote{We continue to use the tensor notation introduced in
Sect.~4.}
\begin{equation}
 {\rm YB}(t)_{klm}~~=~~0~~~,                                    \label{eq:YBE5}
\end{equation}
where, by definition,
\begin{equation}
 {\rm YB}(t)_{klm}~~=~~[ \, t_{kl} \, , \, t_{km} \, ] \; + \;
                       [ \, t_{kl} \, , \, t_{lm} \, ] \; - \;
                       [ \, t_{km} \, , \, t_{ml} \, ]~~~,      \label{eq:YBE6}
\end{equation}
and where $t$ is supposed to be a linear combination of the Casimir tensor $C$
for $\gotg\,$ and some fixed involutive automorphism $\sigma$ of $\gotg\,$,
\begin{equation}
 t~=~aC \, + \, b\;\!\sigma~~~.                                    \label{eq:t}
\end{equation}
Note that, due to the identities (\ref{eq:ID7})--(\ref{eq:ID9}), we have
\begin{eqnarray}
 {\rm YB}(t)_{klm}
 &=& ( \, a_{kl} a_{km} - a_{kl} a_{lm} - a_{km} a_{ml} \, ) \;
     [ \, C_{kl} \, , \, C_{km} \, ]                               \nonumber \\
 & & + \, ( \, a_{kl} b_{km} - a_{kl} b_{lm} - b_{km} b_{ml} \, ) \;
     [ \, C_{kl} \, , \, \sigma_{km} \, ]                          \nonumber \\
 & & - \, ( \, a_{km} b_{kl} - a_{km} b_{ml} - b_{kl} b_{lm} \, ) \;
     [ \, C_{km} \, , \, \sigma_{kl} \, ]                          \nonumber \\
 & & - \, ( \, a_{lm} b_{kl} + a_{ml} b_{km} - b_{kl} b_{km} \, ) \;
     [ \, C_{lm} \, , \, \sigma_{kl} \, ]~~~.                   \label{eq:YBE7}
\end{eqnarray}
As a further input, we must specify what kind of functions of the spectral
parameters we admit for the coefficients $a$ and $b$ in eqn (\ref{eq:t}).

To this end, we first of all set $\; {\cal R}_0 = \field \;$ (the ground field
$\field$ will for simplicity be supposed to be either $\Real$ or $\Complex$)
and, for any integer $n\!\geq\!1$, we let ${\cal R}_n$ denote the ring of
$\field$-valued functions $f$ defined on open, dense domains $D_f$ in
$\field^n$, i.e.,
\[
 {\cal R}_n~~=~~\{ \, (f,D_f) \, |~D_f~\mbox{is open and dense in}~\field^n
                                  \, ,~f: \, D_f \longrightarrow \field~\}~~~.
\]
Of course, addition and multiplication in ${\cal R}_n$ are defined as usual,
i.e., pointwise,
\begin{eqnarray*}
 (f,D_f) + (g,D_g)
 &=& (f|_{D_f\cap D_g} + g|_{D_f\cap D_g} \, , D_f\cap D_g)~~~,              \\
 (f,D_f) \cdot (g,D_g)
 &=& (f|_{D_f\cap D_g} \cdot g|_{D_f\cap D_g} \, , D_f\cap D_g)~~~,
\end{eqnarray*}
which makes sense, because, in any topological space, the intersection of two
open dense subsets is again open and dense. Next, let $(R_n)_{n\geq 0}$ be a
family of subrings $R_n$ of ${\cal R}_n$ containing the constant functions,
with
$\; R_0 = \field \,$, which is closed under division as well as under extension
and restriction of variables, i.e., satisfies the following conditions:
\begin{enumerate}
 \item[(a)] Each $R_n$ is a division ring, i.e., every non-zero element
            $(f,D_f)$ in $R_n$ has an inverse $(f^{-1},D_{f^{-1}})$
            in $R_n$. Explicitly, this means that for every non-zero
            $\, (f,D_f) \smin R_n \,$, \\
            (a$_1$) $D_{f^{-1}} = \{ \lambda \smin D_f | f(\lambda)\!\neq\!0
\}$
                    is open and dense in $\field^n$, \\
            (a$_2$) $(f^{-1},D_{f^{-1}}) \smin R_n$.
 \item[(b)] For $(f,D_f)$ in $R_n$ and $1 \!\leq\! k \!\leq\! n$, the extension
            $(f_{(k)},D_{f_{(k)}})$, defined by
            \vspace{-0.2cm}
            \[
             D_{f_{(k)}}~~=~~\{ \, (\lambda_1,\ldots,\lambda_{n+1}) \smin
                                      \field^{n+1}~|~
            (\lambda_1,\ldots,\lambda_{k-1},
             \lambda_{k+1},\ldots,\lambda_{n+1}) \smin D_f \, \}~~~,
            \vspace{-0.2cm}
            \]
            \[
             f_{(k)}(\lambda_1,\ldots,\lambda_{n+1})~~
             =~~f(\lambda_1,\ldots,\lambda_{k-1},
                  \lambda_{k+1},\ldots,\lambda_{n+1})~~~,
            \vspace{0.1cm}
            \]
            is in $R_{n+1}$; more precisely, this is the extension which is
            constant in the $k$-th variable.
 \item[(c)] For $(f,D_f)$ in $R_n$ and $1 \!\leq\! k \!\leq\! n-1$ and for
            any number $\zeta$ in $\, {\rm pr}_k(D_f) \smsubset \field \,$
            (${\rm pr}_k$ denotes the $k$-th projection from $\field^n$ to
            $\field$: $~{\rm pr}_k(\lambda_1,\ldots,\lambda_n) = \lambda_k
\,$),
            the restriction $(f^{(k)}_\zeta,D_{f^{(k)}_\zeta})$, defined by
            \vspace{-0.2cm}
            \[
             D_{f^{(k)}_\zeta}~~=~~\{ \, (\lambda_1,\ldots,\lambda_{n-1}) \smin
                                          \field^{n-1}~|~
             (\lambda_1,\ldots,\lambda_{k-1},\zeta,
              \lambda_{k+1},\ldots,\lambda_{n-1}) \smin D_f \, \}~~~,
            \vspace{-0.4cm}
            \]
            \[
             f^{(k)}_\zeta(\lambda_1,\ldots,\lambda_{n-1})~~
             =~~f(\lambda_1,\ldots,\lambda_{k-1},\zeta,
                  \lambda_{k+1},\ldots,\lambda_{n-1})~~~,
            \vspace{0.1cm}
            \]
            is in $R_{n-1}$; in particular, this requires $D_{f^{(k)}_\zeta}$
            to be open and dense in $\field^{n-1}$. It should also be noted
            that the set ${\rm pr}_k(D_f)$ of admissible $\zeta$'s is itself
            open and dense in $\field$, because ${\rm pr}_k$ is continuous,
            open and onto.
\end{enumerate}
Important examples for such families are rings of rational functions and rings
of meromorphic functions in several complex variables.

Note that repeated application of the extension property (b) above leads
to various embeddings of $R_m$ into $R_n$, as soon as $m \!\leq\! n$.
In particular, we have the embeddings
\[
 R_1 \longrightarrow R_n~~,~~f \longmapsto f_k~~~
 \Big(~f_k(\lambda_1,\ldots,\lambda_n)~=~f(\lambda_k)~\Big)
\]
for $1 \!\leq\! k \!\leq\! n$ and
\[
 R_2 \longrightarrow R_n~~,~~f \longmapsto f_{kl}~~~
 \Big(~f_{kl}(\lambda_1,\ldots,\lambda_n)~=~f(\lambda_k,\lambda_l)~\Big)
\]
for $1 \!\leq\! k,l \!\leq\! n$. With this notation (which is analogous to
that introduced at the beginning of Sect.~4), we are ready to formulate the
classification theorem:

\begin{theorem} \label{th:A1}
  Let $(R_n)_{n\geq 0}$ be a family of function rings satisfying the conditions
  given above. Let $a$ and $b$ be in $R_2$. Then $\; t = aC + b\:\!\sigma \;$
  satisfies the classical Yang-Baxter equation (\ref{eq:YBE5}) if and only if
  there exist a function $g\smin R_1$ and a non-constant function $f\smin R_1$
  such that $t$ takes one of the following forms:
  \begin{eqnlist}
   t_{kl} &=& g_l \, {C_{kl} \over f_k-f_l}~~~,     \label{CYBEsol:i} \\[0.1cm]
   t_{kl} &=& g_l \, {C_{kl} + \sigma_{kl} \over f_k-f_l}~~~,
                                                   \label{CYBEsol:ii} \\[0.1cm]
   t_{kl} &=& g_l \, \left( {C_{kl} \over f_k-f_l} \; - \;
                            {\sigma_{kl} \over f_k+f_l} \right)~~~.
                                                            \label{CYBEsol:iii}
  \end{eqnlist}
  If in addition $t$ is antisymmetric, it is of the form \ref{CYBEsol:i})
  or \ref{CYBEsol:ii}), with $g$ constant.
\end{theorem}

\begin{proof}
  Looking at eqn (\ref{eq:YBE6}) we see that if $t_{kl}$ is a solution of the
  classical Yang-Baxter equation (\ref{eq:YBE5}), then so is any multiple
  $g_l t_{kl}$. Hence using the identities (\ref{eq:ID7})--(\ref{eq:ID9}),
  it is easily verified that the formulas given above do provide solutions
  of eqn (\ref{eq:YBE5}).

  \medskip
  Conversely, suppose now that $t$ satisfies the classical Yang-Baxter
  equation (\ref{eq:YBE5}). As the four commutators appearing in eqn
  (\ref{eq:YBE7}) are lineary independent, we obtain four equations
  on the coefficients $a$ and $b$:
  \begin{eqnarray}
   a_{kl} a_{km} - a_{kl} a_{lm} - a_{km} a_{ml} &=& 0~~~,  \label{eq:YBE7a} \\
   a_{kl} b_{km} - a_{kl} b_{lm} - b_{km} b_{ml} &=& 0~~~,  \label{eq:YBE7b} \\
   a_{km} b_{kl} - a_{km} b_{ml} - b_{kl} b_{lm} &=& 0~~~,  \label{eq:YBE7c} \\
   a_{lm} b_{kl} + a_{ml} b_{km} - b_{kl} b_{km} &=& 0~~~.     \label{eq:YBE7d}
  \end{eqnarray}
  Now if $a \!=\! 0$, eqns (\ref{eq:YBE7b})--(\ref{eq:YBE7d}) imply
  \mbox{$~b_{kl} b_{km} =  b_{kl} b_{lm} = b_{km} b_{ml} = 0~$} and
  hence $\, b = 0 \,$, because $R_n$, being a division ring, has no
  zero divisors. Thus we get only the trivial solution $\, t = 0$,
  contained in any of the above cases ($g = 0$). We may therefore
  assume without loss of generality that $a \!\neq\! 0$ and divide
  eqn (\ref{eq:YBE7a}) by $\, a_{kl} a_{km} a_{lm} \,$ to obtain
  \begin{equation}  \label{eq:star}
   {1\over a_{km}}~~=~~{1\over a_{lm}}~-~{a_{ml} \over a_{lm}} \,
                                         {1\over a_{kl}}~~~.
  \end{equation}
  The lhs.\ of this equation does not depend on $\lambda_l$. Hence choosing
  a fixed number $\zeta$ in $\; {\rm pr}_1(D_a) \cap {\rm pr}_2(D_a) \;$
  (a set which, as noted in condition (c) above, is the intersection of two
  open dense subsets of $\field$ and hence is again open and dense in $\field$)
  and setting
  \[
   r(\lambda)~=~ a(\zeta,\lambda)~~~,~~~
   s(\lambda)~=~- \, {a(\lambda,\zeta) \over a(\zeta,\lambda)}~~~,~~~
   f(\lambda)~=~{1 \over a(\lambda,\zeta)}~~~,
  \]
  we obtain
  \begin{equation} \label{eq:twostars}
   {1 \over a_{km}}~=~r_m \, + \, s_m f_k~~~~~\mbox{and}~~~~~
   {1 \over a_{kl}}~=~r_l \, + \, s_l f_k~~~.
  \end{equation}
  Inserting this into eqn (\ref{eq:star}) we infer that
  \begin{equation} \label{eq:threestars}
   a_{lm} r_m \, + \, a_{ml} r_l \, - \, 1~
   =~- \, (a_{lm} s_m \, + \, a_{ml} s_l) \, f_k~~~.
  \end{equation}
  Suppose now that $~a_{lm} s_m + a_{ml} s_l \neq 0 \,$. Then dividing eqn
  (\ref{eq:threestars}) by $\; a_{lm} s_m + a_{ml} s_l \,$, we would conclude
  that $f$ must be constant. But $f$ cannot be constant, because otherwise
  $a_{kl}$ would depend on its second argument $\lambda_l$ only, i.e.,
  $\, a_{kl} = r_l \,$ with $\, r \smin R_1 \,$, and as a result, eqn
  (\ref{eq:YBE7a}) would read
  \[
    0~=~r_l r_m \, - \, r_l r_m \, - \, r_m r_l~=~- \, r_m r_l~~~,
  \]
  implying, by the same argument as before (absence of zero divisors)
  that $a$ would have to vanish, contrary to the assumption. Hence
  \mbox{$~a_{lm} s_m + a_{ml} s_l = 0 \,$,} which means that the rescaled
  coefficients $\, \tilde{a}_{lm} = a_{lm} s_m \,$ are antisymmetric:
  \mbox{$\, \tilde{a}_{ml} = - \tilde{a}_{lm} \,$}.

  To summarize, we have shown that from the original solution $t$ of the
  classical Yang-Baxter equation (\ref{eq:YBE5}) of the form (\ref{eq:t}),
  we obtain a new solution of that form, namely
  \begin{equation} \label{eq:fourstars}
   \tilde{t}_{kl} = t_{kl} s_l~~~,~~~\tilde{a}_{kl} = a_{kl} s_l~~~,~~~
   \tilde{b}_{kl} = b_{kl} s_l~~~,
  \end{equation}
  with the additional bonus that this new solution has antisymmetric
  $a$-coefficients:
  \[
   \tilde{a}_{lk} = - \tilde{a}_{kl}~~~.
  \]
  In particular, this means that eqns (\ref{eq:YBE7a})--(\ref{eq:YBE7d})
  are also satisfied when the $a$'s and $b$'s are replaced by the $\tilde{a}$'s
  and $\tilde{b}$'s.

  Now replacing the $a$'s in eqn (\ref{eq:star}) by $\tilde{a}$'s we see that
  \mbox{$~\tilde{a}_{km}^{-1}~=~\tilde{a}_{kl}^{-1} \, - \,
          \tilde{a}_{ml}^{-1} \,$,}
  while inserting eqn (\ref{eq:fourstars}) in eqn (\ref{eq:twostars}) then
gives
  \[
   \tilde{a}_{km}^{-1}~=~\tilde{a}_{kl}^{-1} \, - \, \tilde{a}_{ml}^{-1}~
   =~a_{kl}^{-1} s_l^{-1} \, - \, a_{ml}^{-1} s_l^{-1}~
   =~\left( r_l s_l^{-1} + f_k \right) - \left( r_l s_l^{-1} + f_m \right)~~~,
  \]
  i.e.,
  \begin{equation} \label{eq:takl}
   \tilde{a}_{kl}^{-1}~=~f_k \, - \, f_l~~~.
  \end{equation}
  Defining $\, g = s^{-1} \,$ we get
  \begin{equation} \label{eq:akl}
   a_{kl}~=~{g_l \over f_k-f_l}~~~.
  \end{equation}
  Thus if $b \!=\! 0$, we arrive at the solution (\ref{CYBEsol:i}).
  Suppose therefore that $b \!\neq\! 0$. Dividing eqn (\ref{eq:YBE7d}),
  with $a$'s and $b$'s replaced by $\tilde{a}$'s and $\tilde{b}$'s,
  by $\, \tilde{b}_{kl} \tilde{b}_{km} \tilde{a}_{lm} \,$ gives
  \[
   \tilde{b}_{kl}^{-1}~=~\tilde{b}_{km}^{-1} \, - \, \tilde{a}_{lm}^{-1}
                      ~=~\tilde{b}_{km}^{-1} \, - \, f_k \, + \, f_m~~~,
  \]
  i.e.,
  \[
   \tilde{b}_{kl}^{-1} \, + \, f_l~=~\tilde{b}_{km}^{-1} \, + \, f_m~~~.
  \]
  Both sides of this equation can depend on $\lambda_k$ only (the lhs.\ does
  not depend on $\lambda_m$ and the rhs.\ does not depend on $\lambda_l$),
  so there must exist a function $\, h \smin R_1 \,$ such that
  \begin{equation} \label{eq:tbkl}
   \tilde{b}_{kl}^{-1}~=~h_k \, - \, f_l~~~.
  \end{equation}
  Now dividing eqn (\ref{eq:YBE7c}), with $a$'s and $b$'s replaced by
  $\tilde{a}$'s and $\tilde{b}$'s, by $\, \tilde{a}_{km} \tilde{b}_{kl} \,$
  and inserting eqns (\ref{eq:takl}) and (\ref{eq:tbkl}), we get
  \[
   {h_k-f_l \over h_m-f_l}~+~{f_k-f_m \over h_l-f_m}~~=~~1~~~,
  \]
  or
  \begin{equation} \label{eq:fivestars}
   {h_l-f_k \over f_l-h_k}~~=~~{h_l-f_m \over f_l-h_m}~~~.
  \end{equation}
  Again, both sides of this last equation can depend on $\lambda_l$ only, so
  choosing two fixed numbers $\zeta$ and $\zeta'$ in $\, D_{f}\cap D_{h} \,$
  and setting $~~\mbox{$\alpha = f(\zeta)$}~,~\mbox{$\beta = h(\zeta)$}~,~
  \mbox{$\alpha' = f(\zeta')$}~,~\mbox{$\beta' = h(\zeta')$}~$, we see that
  eqn (\ref{eq:fivestars}), with $~~\mbox{$\lambda_l \!=\! \lambda$}~,~
  \mbox{$\lambda_k \!=\! \zeta$}~,~\mbox{$\lambda_m \!=\! \zeta'$}~$, becomes
  \[
   (\beta-\beta') \, h_l~=~(\alpha-\alpha') \, f_l \, + \,
                            \alpha'\beta \, - \, \alpha\beta'~~~.
  \]
  As $f$ is a non-constant function, we can choose $\zeta$ and $\zeta'$ in
  such a way that $\alpha \!\neq\! \alpha'$. But then, for the same reason,
  $\beta \!\neq\! \beta'$, and hence $h$ is a non-constant function given by
  \begin{equation} \label{eq:sixstars}
    h~=~\gamma f \, + \, \delta~~~,
  \end{equation}
  with
  \begin{equation} \label{eq:sevenstars}
    \gamma~=~{\alpha-\alpha' \over \beta-\beta'}~\neq~0~~~~~\mbox{and}~~~~~
    \delta~=~{\alpha'\beta-\alpha\beta' \over \beta-\beta'}~~~.
  \end{equation}
  On the other hand, eqn (\ref{eq:sixstars}) taken at $\lambda \!=\! \zeta$
  and at $\lambda \!=\! \zeta'$ gives
  \begin{equation} \label{eq:eightstars}
   \beta~=~\gamma\alpha \, + \, \delta~~~~~\mbox{and}~~~~~
   \beta'~=~\gamma\alpha' \, + \, \delta~~~.
  \end{equation}
  Eqns (\ref{eq:sevenstars}) and (\ref{eq:eightstars}) imply
  \[
   \gamma~=~\pm \, 1~~~~~\mbox{and}~~~~~\delta \, (\gamma + 1)~=~0
  \]
  and hence either
  \[
   h~=~\pm \, f~~~,
  \]
  or
  \[
   h~=~- \, f \, + \, \delta~~~.
  \]
  The first case yields
  \begin{equation} \label{eq:bkl1}
   b_{kl}~=~\pm \, {g_l \over f_k \mp f_l}~~~,
  \end{equation}
  i.e., solutions (\ref{CYBEsol:ii}) and (\ref{CYBEsol:iii}), whereas the
  second case results in
  \begin{equation} \label{eq:bkl2}
   b_{kl}~=~{g_l \over \delta-f_k-f_l}~~~.
  \end{equation}
  So shifting $f$ by $\delta/2$ gives again (\ref{CYBEsol:iii}),
\end{proof}

\subsection{Proof of the No-Go Theorem}

Here we shall prove the no-go theorem given at the end of Sect.~4:

\begin{proof}
  Write the quantities $a_{kl}$ and $b_{kl}$ in eqn (\ref{eq:PB4c}) as linear
  combinations of $C_{kl}$ and $\sigma_{kl}$, i.e.,
  \begin{eqnarray*}
   a_{kl} &=& f_{kl} \, C_{kl} \, + \, g_{kl} \, \sigma_{kl}~~~,      \\[0.1cm]
   b_{kl} &=& \phi_{kl} \, C_{kl} \, + \, \gamma_{kl} \, \sigma_{kl}~~~,
  \end{eqnarray*}
  where the functions $f_{kl}$, $g_{kl}$, $\phi_{kl}$ and $\gamma_{kl}$
  depend on $\lambda_k$ and $\lambda_l$ and are antisymmetric.
  Now looking at the expressions in curly brackets in eqn (\ref{eq:PB4c}),
  we see that the terms containing a commutator with $C_{kl} + C_{km}$
  (plus cyclic permutations) give rise to $C-\sigma$ type commutators but
  not to $C-C$ type commutators: the only contribution to the latter comes
  from the Yang-Baxter term. Thus according to eqn (\ref{eq:YBE7a}), we get
  \begin{eqnarray*}
   f_{kl} \, f_{km} \, - \, f_{kl} \, f_{lm} \, - \, f_{km} \, f_{ml}~=~0~~~,\\
   \phi_{kl} \, \phi_{km} \, - \, \phi_{kl} \, \phi_{lm} \, - \,
                                  \phi_{km} \, \phi_{ml}~=~0~~~.
  \end{eqnarray*}
  Arguing as in the proof of Theorem \ref{th:A1} to derive eqn (\ref{eq:takl})
  from eqn (\ref{eq:YBE7a}), we conclude that there exist non-constant
  functions $f$ and $\phi$ of one real variable $\lambda$ and real numbers
  $\epsilon_1$ and $\epsilon_2$ which are either $0$ or $1$, such that
  \begin{eqnarray*}
   f_{kl} &=& {\epsilon_1 \over f_k-f_l}~~~,                                 \\
   \phi_{kl} &=& {\epsilon_2 \over \phi_k-\phi_l}~~~.
  \end{eqnarray*}
  On the other hand, using (\ref{eq:d2}), the $C_{kl}$-coefficients of $a_{kl}$
  and $b_{kl}$ can be computed in terms of $\alpha_{kl}$, whence we have the
  following two equations:
  \begin{eqnarray*}
   {\epsilon_1 \over f_k-f_l}
   &=& - \, {1\over \lambda_k-\lambda_l}
       \left( {2\lambda_k^2 \over 1\!-\!\lambda_k^2} \, \alpha_{kl} \, + \,
              {2\lambda_l^2 \over 1\!-\!\lambda_l^2} \, \alpha_{lk}
       \right)~~~,                                                           \\
   {\epsilon_2 \over \phi_k-\phi_l}
   &=& {1 \over \lambda_k-\lambda_l}
       \left( {2\lambda_k^2 \over 1\!-\!\lambda_k^2} \, (\alpha_{kl}-1) \, + \,
              {2\lambda_l^2 \over 1\!-\!\lambda_l^2} \, (\alpha_{lk}-1)
       \right)~~~.
  \end{eqnarray*}
  Adding the two equations gives the following $\alpha$-independent
  consistency condition:
  \[
   {\epsilon_1 \over f_k-f_l} \, + \, {\epsilon_2 \over \phi_k-\phi_l}~~
   =~~- \, {1 \over \lambda_k-\lambda_l}
      \left( {2\lambda_k^2 \over 1\!-\!\lambda_k^2} \, + \,
             {2\lambda_l^2 \over 1\!-\!\lambda_l^2} \right)~~~.
  \]
  Now we use the following singularity argument\footnote{We are indebted to
  C.~Nowak for suggesting this simple reasoning.}. For $\lambda_k\!=\!\pm 1$,
  the rhs.\ of this equation is singular for all values of $\lambda_l$, while
  the denominators on the lhs.\ can vanish only for certain values of
  $\lambda_l$ because the functions $f$ and $\phi$ are not constant.
  Therefore this last equation cannot be satisfied, and a contradiction
  is established,
\end{proof}

{\bf Acknowledgments}: We would like to thank C.~Nowak and U.~Pinkall for
helpful discussions and pointing out ref.~\cite{BFPP}.



\begin{thebibliography}{99}

\bibitem{AvTa} J.~Avan and M.~Talon: {\em Rational and Trigonometric Constant
 Non-Antisymmetric R-Matrices},
 Phys. Lett.~{\bf B~241} (1990) 77-82.

\bibitem{BaVi} O.~Babelon and C.-M.~Viallet: {\em Hamiltonian Structures and
 Lax Equations}, Phys.~Lett.~{\bf 237~B} (1990) 411-416.

\bibitem{BFPP} F.E.~Burstall, D.~Ferus, F.~Pedit and U.~Pinkall:
 {\em Harmonic Tori in Symmetric Spaces and Commuting Hamiltonian Systems
 on Loop Algebras}, University of Bath / Technische Universit\"at Berlin /
 Emory University Atlanta preprint, 1991.

\bibitem{DEM} H.J.~de Vega, H.~Eichenherr and J.-M.~Maillet: {\em Classical
 and Quantum Algebras of Non-Local Charges in Sigma Models},
 Commun.~Math.~Phys.~{\bf~92} (1984) 507-524.

\bibitem{EF} H.~Eichenherr and M.~Forger: \newline
 {\em On the Dual Symmetry of the Nonlinear Sigma Models},
 Nucl.~Phys.~{\bf B~155} (1979) 381-393. \newline
 {\em More about Nonlinear Sigma Models on Symmetric Spaces},
 Nucl.~Phys.~{\bf B~164} (1980) 528-535 \& {\bf B~282} (1987) 745-746
 (erratum). \newline
 {\em Higher Local Conservation Laws for Nonlinear Sigma Models on Symmetric
 Spaces}, Commun.~Math.~Phys.~{\bf 82} (1981) 227-255.

\bibitem{FaTa} L.D.~Faddeev and L.A.~Takhtajan: {\em Hamiltonian Methods in the
 Theory of Solitons}, Springer-Verlag, Berlin 1987.

\bibitem{For} M.~Forger: {\em Nonlinear Sigma Models on Symmetric Spaces}.
In{\em Nonlinear Partial Differential Operators and Quantization Procedures},
 proceedings, Clausthal, Germany 1981, eds: S.I. Andersson and H.D. Doebner;
 Lecture Notes in Mathematics {\bf 1037}, Springer-Verlag, Berlin 1983.

\bibitem{FLS} M.~Forger, J.~Laartz and U.~Sch\"aper: {\em Current Algebra of
 Classical Non-Linear Sigma Models}, Freiburg University preprint THEP 91/10,
 to be published in Commun.~Math.~Phys..

\bibitem{FrMa} L.~Freidel and J.M.~Maillet: \newline
 {\em Quadratic algebras and integrable systems}, preprint LPTHE-24/91,
 \newline
 {\em On Classical and Quantum Integrable Field Theories Associated to
 Kac-Moody Current Algebras}, preprint LPTHE-25/91.

\bibitem{Hel} S.~Helgason: {\em Differential Geometry, Lie Groups, and
 Symmetric Spaces}, Academic Press, New York 1978

\bibitem{La} J.~Laartz: {\em The Extension Structure of 2D Massive Current
 Algebras}, Freiburg University preprint THEP 91/21,

\bibitem{Mai2} J.-M.~Maillet: {\em Kac-Moody Algebra and Extended Yang-Baxter
 Relations in the $O(N)$ Non-Linear Sigma Model},
 Phys.~Lett.~{\bf 162~B} (1985) 137-142.

\bibitem{Mai3} J.-M.~Maillet: {\em Hamiltonian Structures for Integrable
 Classical Theories from Graded Kac-Moody Algebras},
 Phys.~Lett.~{\bf 167~B} (1986) 401-405.

\bibitem{Mai1} J.-M.~Maillet: {\em New Integrable Canonical Structures in
 Two-Dimensional Models}, Nucl.~Phys.~{\bf B~269} (1986) 54-76.

\end{thebibliography}
\end{document}